\documentclass[review]{elsarticle}

\usepackage{lineno,hyperref}
\usepackage{amssymb}
\usepackage{floatrow}
\usepackage{amsmath,amsfonts}
\usepackage{array}
\usepackage{textcomp}
\usepackage{stfloats}
\usepackage{url}
\usepackage{verbatim}
\usepackage{graphicx}
\usepackage{subcaption}
\usepackage{tikz}
\usetikzlibrary{shadows.blur}
\usetikzlibrary{shapes.symbols}
\usetikzlibrary{shapes.geometric, arrows}
\usepackage[ruled,vlined,titlenumbered]{algorithm2e}
\usepackage{import}
\usepackage{float}
\usepackage{adjustbox}

\usepackage{xparse}%
\usepackage{blindtext}
\usepackage{morewrites}
\usepackage{wrapfig}
\usepackage{xkeyval}%
\usepackage{algorithmic}
\usepackage{pifont}
\usepackage{multirow}
\usepackage[tableposition=top]{caption}
\usepackage{soul}
\usepackage{xcolor}

\newcommand{\cmark}{\ding{51}}%
\newcommand{\xmark}{\ding{55}}%

\newwrite\authorbibfile%

\AtBeginDocument{%
  \immediate\openout\authorbibfile=\jobname.aub%
}%

\AtEndDocument{%
\immediate\closeout\authorbibfile
\InputIfFileExists{\jobname.aub}{}{}
}%

\makeatletter

\define@key{authorbib}{scale}[1]{%
\def\AuthorbibKVMacroScale{#1}%
}

\define@key{authorbib}{wraplines}[10]{%
\def\AuthorbibKVMacroWraplines{#1}%
}

\define@key{authorbib}{imagewidth}[4cm]{%
\def\AuthorbibKVMacroImagewidth{#1}%
}

\define@key{authorbib}{overhang}[10pt]{%
\def\AuthorbibKVMacroOverhang{#1}%
}

\define@key{authorbib}{imagepos}[l]{%
\def\AuthorbibKVMacroImagepos{#1}%
}

\makeatother

\presetkeys{authorbib}{imagepos=l,imagewidth=4cm,wraplines=15,overhang=20pt}{}

\newlength{\AuthorbibTopSkip}
\newlength{\AuthorbibBottomSkip}
\NewDocumentCommand{\authorbibliography}{+o+m+m+m}{%
  \IfNoValueTF{#1}{%
  }{%
    \setkeys{authorbib}{#1}%
    \immediate\write\authorbibfile{%
      \string\begin{wrapfigure}[\AuthorbibKVMacroWraplines]{\AuthorbibKVMacroImagepos}[\AuthorbibKVMacroOverhang]{\AuthorbibKVMacroImagewidth}^^J
        \string\includegraphics[scale=\AuthorbibKVMacroScale]{#2}^^J
        \string\end{wrapfigure}^^J
    }%
  }%
  \IfNoValueTF{#3}{%
    \typeout{Warning: No author name}%
  }{%
    \immediate\write\authorbibfile{%
      \unexpanded{\vspace{\AuthorbibTopSkip}}^^J
      \string\noindent\relax
      \unexpanded{\textbf{#3}\par}^^J
      \string\noindent\relax
      \unexpanded{#4}^^J%
      \unexpanded{\vspace{\AuthorbibBottomSkip}}^^J
      }%
  }%
}%

\modulolinenumbers[5]

\journal{Journal}

\begin{document}

\begin{frontmatter}

\title{ModularFed: Leveraging Modularity in Federated Learning Frameworks}

\author[label1]{Mohamad Arafeh}
\author[label2]{Hadi Otrok}
\author[label5]{Hakima Ould-Slimane}
\author[label3,label4]{Azzam Mourad}
\author[label1]{Chamseddine Talhi}
\author[label2]{Ernesto Damiani}

\address[label1]{Department of Software Engineering, Ecole de Technologie Superieure (ETS), Montreal, QC, Canada}
\address[label5]{Department of mathematics and computer science, Universite de Quebec a Trois-Rivieres (UQTR), Canada}
\address[label2]{Center of Cyber-Physical Systems (C2PS), Department of EECS, Khalifa University, Abu Dhabi, UAE}
\address[label3]{Cyber Security Systems and Applied AI Research Center, Department of CSM, Lebanese American University, Lebanon}
\address[label4]{Division of Science, New York University, Abu Dhabi, United Arab Emirates}



\begin{abstract}
Numerous research recently proposed integrating Federated Learning (FL) to address the privacy concerns of using machine learning in privacy-sensitive firms. However, the standards of the available frameworks can no longer sustain the rapid advancement and hinder the integration of FL solutions, which can be prominent in advancing the field. In this paper, we propose ModularFed, a research-focused framework that addresses the complexity of FL implementations and the lack of adaptability and extendability in the available frameworks. We provide a comprehensive architecture that assists FL approaches through well-defined protocols to cover three dominant FL paradigms: adaptable workflow, datasets distribution, and third-party application support. Within this architecture, protocols are blueprints that strictly define the framework's components' design, contribute to its flexibility, and strengthen its infrastructure. Further, our protocols aim to enable modularity in FL, supporting third-party plug-and-play architecture and dynamic simulators coupled with major built-in data distributors in the field. Additionally, the framework support wrapping multiple approaches in a single environment to enable consistent replication of FL issues such as clients' deficiency, data distribution, and network latency, which entails a fair comparison of techniques outlying FL technologies. In our evaluation, we examine the applicability of our framework addressing three major FL domains, including statistical distribution and modular-based approaches for resource monitoring and client selection. 
\end{abstract}


\end{frontmatter}


\section{Introduction}
Federated learning is a distributed technique that emerged from the need for an architecture to address the increasing restriction on users' data privacy. In FL, users train ML models locally on their devices and send them to external servers while keeping their raw data intact on their devices. Furthermore, the server, in its turn, performs a particular aggregation algorithm, which is the core of any FL approach. Aggregation algorithms have the role of combining the clients' received models into one global model. Therefore, external parties can benefit from users' collected data without violating their privacy.

Despite its promising rationale and the numerous recent investigations \cite{am1, am2, am3, am6}, FL's trustworthiness and performance are majorly affected by statistical, security and resource concerns, restricting further development in the field. For instance, FL implies individual training from clients using their local datasets, which in most cases, is deemed biased toward the clients' behaviours and surroundings. Such a situation signifies a non-identical and independent (Non-IID) data distribution in which data imbalance exists by either size or content (features/classes). Referring to \cite{google}, experiments with Non-IID constitute evidence of serious drawbacks on accuracy and loss developments. Furthermore, the absence of raw data on the server precludes the cleaning and filtering procedures in the traditional Non-FL server that takes place in a centralized learning context. The eminent lack of raw data access pushes researchers to propose alternative resolutions to address the statistical problem. Approaches such as \cite{cluster1, cluster2, cluster3, cluster4} attempted to exploit the connection between the clients and their model parameters by clustering algorithms to separate them into distinct FL contexts. Having only similar clients working together can simulate an IID environment. On the other hand, the works presented in \cite{agg1, agg2, agg3} take advantage of the aggregation procedure to fix the model shift caused by training on Non-IID clients. Another form of solution operates on the starting parameters. For instance, in \cite{arm, am4}, the experiments show that starting from a pre-trained model can boost the performance in a Non-IID environment. In terms of security, various defence mechanisms aim to protect against manipulation \cite{psn1,psn2,security1}, trustworthiness \cite{trust1,trust2,trust3}, and inference attacks \cite{infer1,infer2}. Regarding resource consumption, development focuses on reducing the communication overhead as in \cite{comm1,comm2} or energy consumption as in \cite{cons1, cons2, cons3, cons4}. Other approaches in \cite{selection3,rl} acted on the produced clients' model quality to reduce the communication rounds. Their experiments show a significant improvement in weight development through selection algorithms, increasing the convergence rate and reducing the total cost. 

From the aforementioned approaches, we underline essential facts which are the motivation behind this work: 
\begin{itemize}
    \item The majority of the aforementioned approaches focuses on three areas: aggregation, client selection, communication \cite{agg1,agg2,selection2,rl,comm1,comm2, am5} that can be targeted to create a flexible environment for future solutions.
    \item With few modifications, it is possible to integrate existing solutions from other domains into FL to improve it. For instance, the involvement of blockchain in FL improves the underlying communication, such as robustness and availability \cite{bch1,bch2,bch3}.
    \item Integrating FL increases the complexity, which scales with the needs of the various mechanisms such as parallelism, data distribution, client selection, varying training engine, model analysis, data management, bandwidth and resource allocation.
    \item The currently available frameworks are limited in terms of standardization and customizability, interfering with the search for new solutions to the said challenges.  
\end{itemize}


Therefore, essential details should be considered when working in a complex environment. A framework can solve these problems, but it must have the right level of abstraction; otherwise, it will not be advantageous. To this end, we propose ModularFed, a research-heavy federated learning framework in which we aim to provide a complete set of extendable tools capable of easing the integration of applications in the FL domain, reducing the implementation times while withholding future projects' extendability and scalability. 

In summary, our main contributions are the following:

\begin{itemize}
    \item A modular-based framework supporting modular swapping opens the opportunities for work conjunction or component-specific FL approach.
    \item An intuitive subscription base architecture for seamless third-parties integration.
    \item Central data control supporting external dataset and dynamic distributions simulations.
\end{itemize}

Following the proposed architectures and addressing the aforementioned issues, we provide an exhaustive framework to orchestrate FL approaches and authorize interchanging components' capabilities. As such, we open the opportunities for collaboration between multiple techniques while additionally easing the integration of new methodologies. Our experimental results demonstrate our framework's capabilities in supporting various contextual FL problems and scenarios, such as data distribution diversity, client performance, and resource limitation.  


\section{Related Work}
Various frameworks addressing the integration complexity have been proposed. For instance, TensorFlow Federated \cite{tensorflow}, by Google, the original creator of FL \cite{google}, aims to enhance their TensorFlow engine to support distributed learning. The framework supports essential features such as a subscription to a limited set of events to monitor the execution states while making thirds party integration possible. Additionally, the framework incorporates the commonly used datasets and the necessary tools for an effortless start. However, the framework only supports TensorFlow models, making it harder for researchers to integrate FL with models built using another training engine. Another engine-specific FL framework exists, such as FedToch \cite{fedtorch}, which only supports PyTorch models.

Contrary to the mentioned frameworks, FedML authors, in \cite{fedml}, provide an exhaustive framework that started as a fair benchmarking tool for approaches using federated learning. It integrates a complete federated workflow from client selection, aggregation, and validation. Unlike other approaches, where the tests run only locally, their framework supports message-passing interface-based (MPI) settings, which is the key for parallel client simulation in which procedure execution is within or beyond a single host. Nevertheless, enterprise solutions that embrace the concept of FL, such as Federated AI Technology Enabler (FATE) \cite{fate} and IBM Federated Learning \cite{ibm}, exist. For instance, FATE provides the required mechanism, from secure computation protocols based on homomorphic encryption and multi-party computation (MPC), for industries to integrate FL in their framework. However, the lack of the necessary benchmarking tools, the complex integration, and the limited extensibility makes it harder for researchers to integrate industrial-based frameworks for their works.

\begin{table*}[ht]
	\centering
	\normalsize
	\def\arraystretch{1.5}%
    \caption{Comparison of the supported features between the available frameworks and ours}
	\resizebox{\textwidth}{!}{
    	\begin{tabular}{|l|c|c|c|c|c|c|c|}
    		\hline
    		\multicolumn{1}{|c|}{\textbf{}}          & \textbf{TFF\cite{tensorflow}}    & \textbf{FedTorch\cite{fedtorch}} & \textbf{PySyft} & \textbf{FATE \cite{fate}}   & \textbf{IBM \cite{ibm}}    & \textbf{FedML\cite{fedml}}  & \textbf{Ours}   \\ \hline
    		\textbf{Local   Execution}               & \textbf{\cmark} & \textbf{\cmark}   & \textbf{\cmark} & \textbf{\cmark} & \textbf{\cmark} & \textbf{\cmark} & \textbf{\cmark} \\ \hline
    		\textbf{Distributed   Execution}         & \textbf{\cmark} & \textbf{\cmark}   & \textbf{\cmark} & \textbf{\cmark} & \textbf{\cmark} & \textbf{\cmark} & \textbf{\cmark} \\ \hline
    		\textbf{Virtualization/Real   Clients}   & \textbf{\xmark} & \textbf{\cmark}   & \textbf{\cmark} & \textbf{\cmark}      & \textbf{\cmark}      & \textbf{\cmark} & \textbf{\cmark} \\ \hline
    		\textbf{Supported   Messaging Protocols} & \textbf{-}      & \textbf{MPI}      & \textbf{-}      & \textbf{REST}   & \textbf{-}      & \textbf{MPI}    & \textbf{Any}    \\ \hline
    		\textbf{Secure}           & \textbf{\cmark} & \textbf{\xmark}   & \textbf{\cmark} & \textbf{\cmark} & \textbf{\cmark} & \textbf{\cmark} & \textbf{Any}    \\ \hline
    		\textbf{Modular   Components}            & \textbf{\xmark} & \textbf{\xmark}   & \textbf{\xmark} & \textbf{\xmark} & \textbf{\xmark} & \textbf{\xmark} & \textbf{\cmark} \\ \hline
    		\textbf{Built-In   Monitoring Tools}     & \textbf{\cmark} & \textbf{\xmark}   & \textbf{\xmark} & \textbf{\xmark} & \textbf{\xmark} & \textbf{\xmark} & \textbf{\cmark} \\ \hline
    		\textbf{Third Party   Support}           & \textbf{\cmark} & \textbf{\xmark}   & \textbf{\xmark} & \textbf{\xmark} & \textbf{\xmark} & \textbf{\xmark} & \textbf{\cmark} \\ \hline
    		\textbf{Customizable   Data Center}      & \textbf{\xmark} & \textbf{\xmark}   & \textbf{\xmark} & \textbf{\xmark} & \textbf{\xmark} & \textbf{\xmark} & \textbf{\cmark} \\ \hline
    		\textbf{Diverse Data   Distributor}      & \textbf{\xmark} & \textbf{\xmark}   & \textbf{\xmark} & \textbf{\xmark} & \textbf{\xmark} & \textbf{\xmark} & \textbf{\cmark} \\ \hline
    		\textbf{Client   Simulation}             & \textbf{\xmark} & \textbf{\xmark}   & \textbf{\xmark} & \textbf{\xmark} & \textbf{\xmark} & \textbf{\cmark} & \textbf{\cmark} \\ \hline
    	\end{tabular}  
    }
	\label{tab:comp}
\end{table*}
Table \ref{tab:comp} provides a detailed comparison between our approach and the current federated frameworks. We emphasize the limitation of the current approaches, which we aim to address using ours:

\textbf{Lack of modular component support:} Modularity, in our case, refers to frameworks' supporting individual building blocks with a separate role assigned for each. Together, modular components form a framework that endorses modular interchange. Due to the lack of support for such a concept, the majority tend to build their own environment instead of working within contextual frameworks with a high learning curve while risking the possibility of not supporting their works. Consequently, the experimental results might not demonstrate a fair comparison in terms of distinct implementation differences and unmanaged statistical environments. Addressing these issues, we pursue a layered architecture in our framework called the federated abstract layer (FAL). With FAL, we design our components based on carefully crafted protocols while delegating the behaviours to the following layers. Accordingly, protocols are blueprints that define the interaction between the architecture components, allowing researchers to develop flexible and extensible FL approaches. FAL enables FL to support diverse execution mechanisms, such as distributed execution in virtual containers, real devices, or local parallel execution on the same host. Additionally, FAL facilitates extending approaches with supporting features such as security and bandwidth optimization through model compression modules.  

\textbf{Lack of data management protocols:} Current frameworks only support a few well-known datasets used in FL. Built-in tools back up these datasets for easier management, such as dynamic allocation from the cloud or data distribution simulators. However, they are limited to only the supported datasets, while incorporating new ones compels designing these tools from scratch. In this regard, we integrate the data management as part of the framework and outsource the required protocols and APIs to allow further data expansion depending on the designer's needs. 

\textbf{Third Party Support:} By third parties, we refer to any components with a role not related directly to FL workflow. For instance, external monitoring components such as Tensor-Board and Wandb are considered under this category. Additionally, components built to enhance FL architecture are considered a third party, such as model caching, models' parameters analysis, logging, and others. These tools do not affect or are presented in the federated workflow. However, including them increases the framework's complexity. Thus, frameworks either do not support third-party tools or only support a few well-known ones. In our case, we started our framework design with the concept of supporting third parties by incorporating an observable pattern for our components. As such, a list of third parties, called subscribers, can be attached to any federated application and receive live updates about the state of the execution employed, extending it with further applications such as monitoring, logging, or analysis.

\section{Federated Learning Overview} \label{sec:fl-workflow}
FL is a branch of distributed learning where multiple individual parties collaborate to train a single model. However, unlike distributed learning, FL operates without sharing private data with collaborators. Instead, it invites them to train a model locally on their devices using the same configuration and share only the trained parameters. The following properties define Federated Learning: 
\begin{itemize}
    \item \textbf{Global Model} (GM): Initialized by the server and holds its configuration and the initial parameters values. Preceding any further actions, each working client holds the latest copy of the GM shared by the server.
    \item \textbf{Aggregation:} A procedure the server executes when it receives the required updates from its clients. The received updates are combined and committed to the Global Model at the end of the procedure.
    \item \textbf{Client Selection:} A typical FL context comprises a large pool of clients, which demands considerable computational resources and causes network congestion on the server. Additionally, it is impracticable to guarantee the availability of all the clients in every round, especially when a task might take a considerable amount of time. Thus, most FL approaches incorporate a client selection phase to filter the clients either randomly or by specified constraints \cite{selection1, selection2, selection3, arm}, and only the selected clients will participate in the specified round.
\end{itemize}

\begin{figure}[ht]
\centering
\includegraphics[width=0.78\textwidth]{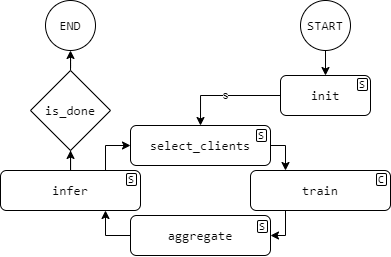}
\caption{Federated Learning Workflow}
\label{fig:fl-workflow}
\end{figure}

In Figure \ref{fig:fl-workflow}, we present an overview of a standard FL workflow. The server starts with the initialization, validating the parameters and issuing the first copy of the Global Model. We follow the same initialization strategy as \cite{google}, where the server starts by initializing the model weights and sharing them with the selected clients. Having clients start from the same initialization improves the loss reduction after weights aggregation. A federated round begins by selecting a set of clients from the available ones. The selected clients receive a copy of the Global Model and proceed with the model training from their local datasets. When the selected clients finish, each sends back the models to the server. The server aggregates the received updates and generates a new, evolved Global Model, substituting the old one. When working in a research environment, part of the dataset is kept to monitor the model evolution and validate the approach's applicability. After each aggregation, the server infers the model metrics from the test dataset and logs them. Eventually, FL stops when reaching the stopping criteria, such as completing a specified number of rounds or achieving a predefined accuracy or loss.

\section{Federated Learning Framework}
\begin{figure*}[ht]
\centering
\includegraphics[width=0.98\textwidth]{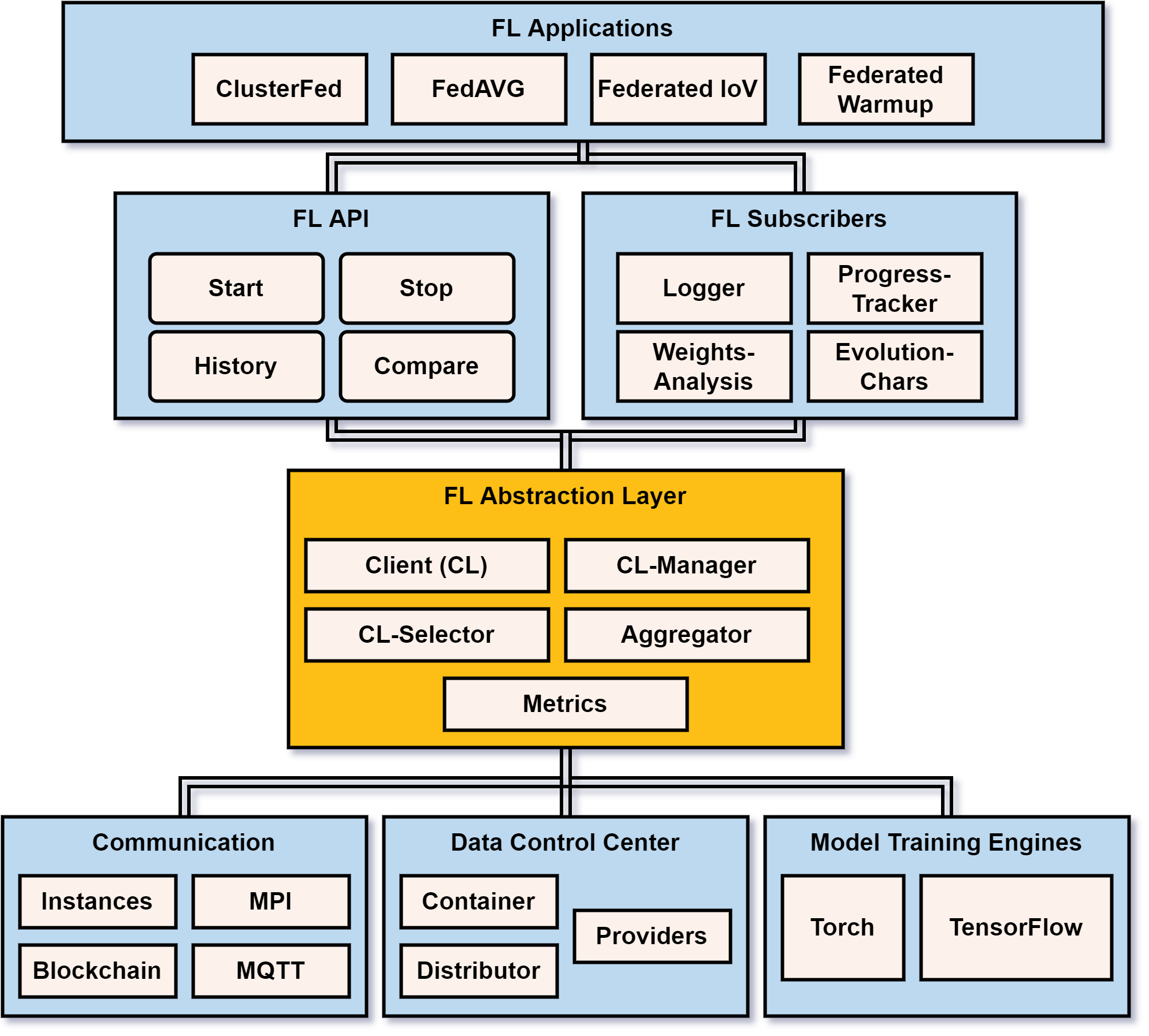}
\caption{Proposed Layered Architecture}
\label{fig:framework}
\end{figure*}

Figure \ref{fig:framework} provides an overview of the proposed framework architecture. Three key components define our architecture from the rest: FL Abstraction Layer (FAL), FL Subscribers, and Data Control Center.

\subsection{FL Abstraction Layer}
\begin{figure}[ht]
\centering
\includegraphics[width=0.78\textwidth]{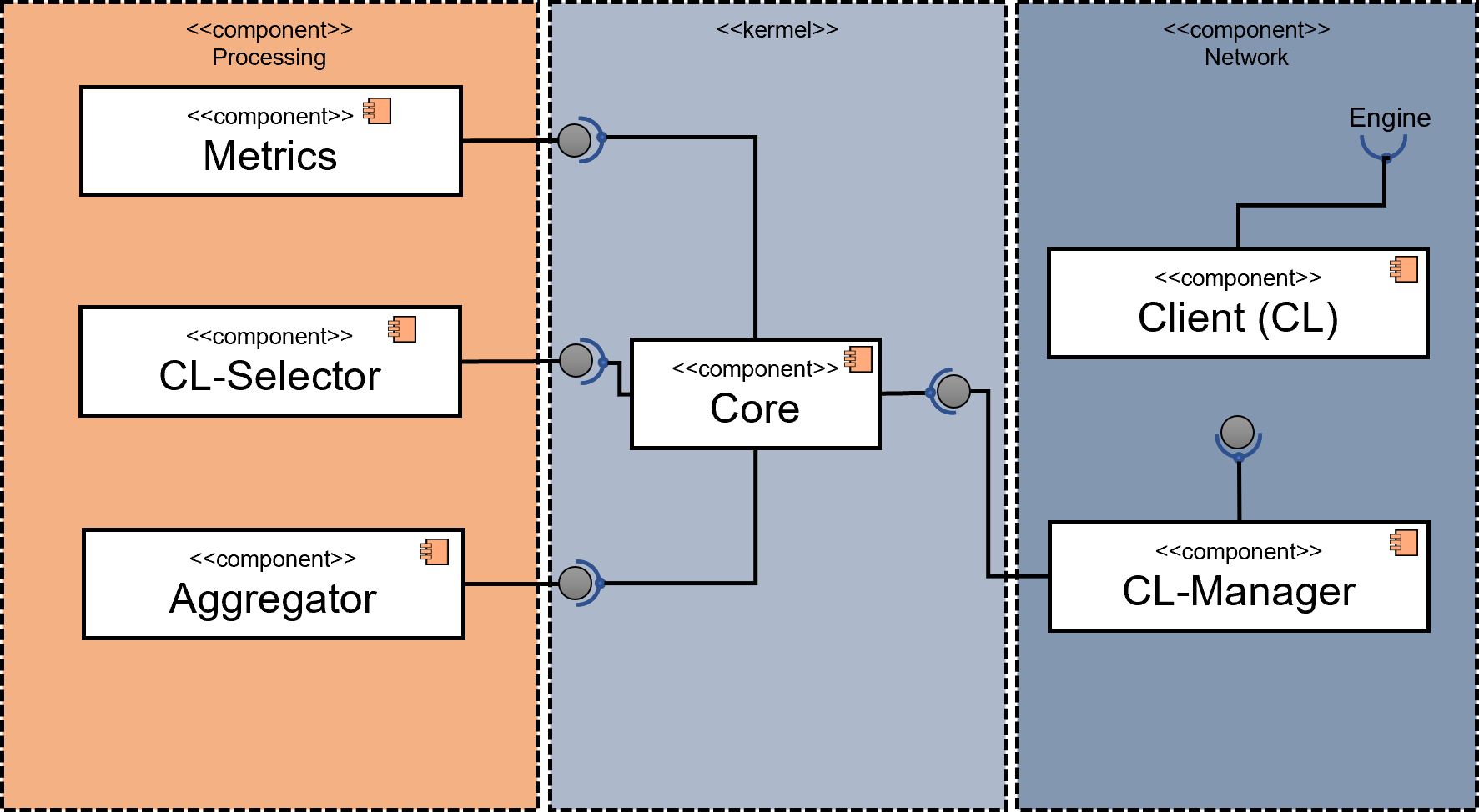}
\caption{FAL UML Components Diagram}
\label{fig:uml}
\end{figure}

\begin{figure}[ht]
\centering
\includegraphics[width=0.75\textwidth]{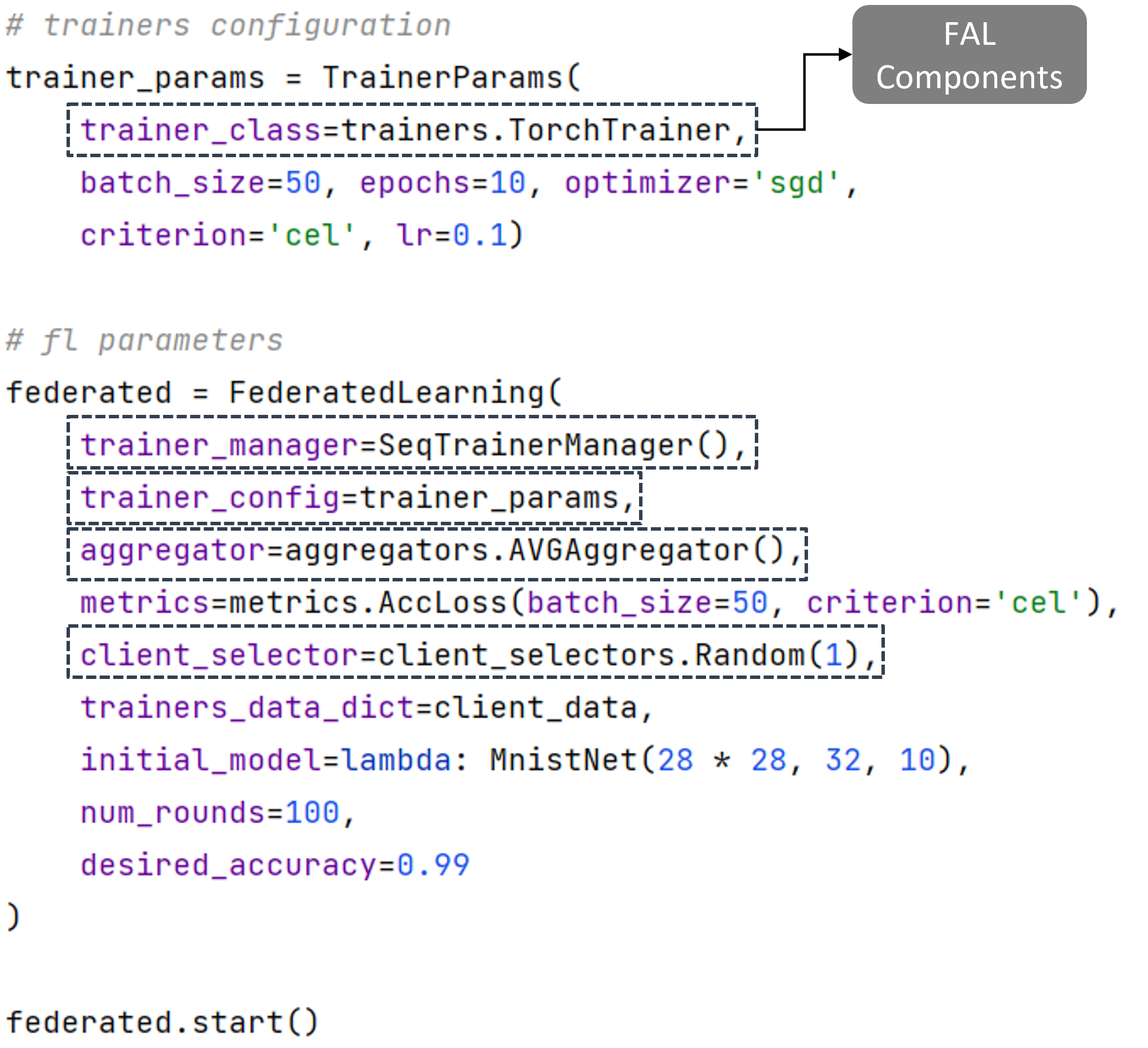}
\caption{FAL Design Pattern}
\label{fig:dep}
\end{figure}

We follow a hierarchical mechanism to qualify extensions and new approaches control over the necessary component dynamically and with minimum modification to the rest of the framework. We standardize FL components through FAL as protocols defining compulsory properties and parameters while delegating the behaviour to the subsequent layers. Federated learning workflow reflects on FAL components and is separated into two higher-order components: Network Components to handle the communication and Processing Components for vital calculations such as aggregation and client selection. Moreover, the kernel is the glue that manages the FAL and advances with the ordered execution of the FL workflow while taking advantage of the FAL flexibility. Figure \ref{fig:uml} presents UML Component Diagram of the FAL, while Figure \ref{fig:dep} shows the entire training implementation with the FAL components highlighted.

The Client component is an interface that enables the framework to simulate diverse training contexts allowing it to adapt to the needed scenarios. For instance, a client simulation can be straightforward, such as a PyTorch instance for model training, or more complex, such as a virtual node or a tangible IoT device. We consider the client component as a separate entity from the core module, which opens opportunities for more complex topologies where clients are free from core control while withholding only the minimum specification posed by FAL architecture. 

\begin{figure}[ht]
\centering
\includegraphics[width=0.65\textwidth]{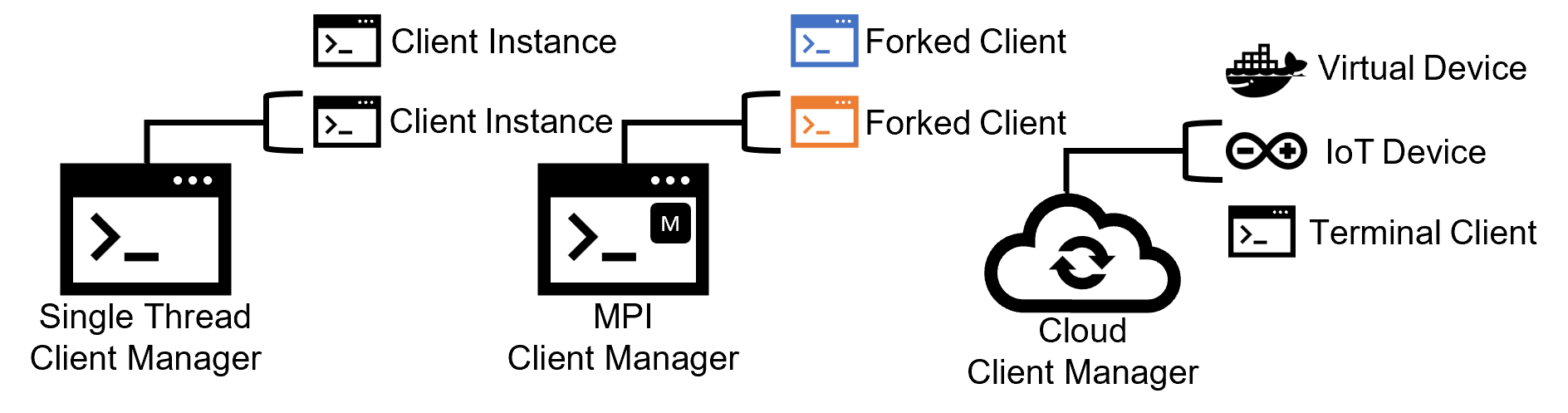}
\caption{The relation between network manager and the supported clients' types}
\label{fig:cl-man}
\end{figure}

On the other hand, CL-Manager is the middleware between the core and clients, responsible for managing the clients, communicating tasks, and global model updates. CL-Manager shapes the interaction between the core and the clients authorizing various behaviours such as parallelism, secure integrated network, or further network optimization such as model compression or communication over any available network protocol. Depending on the case, each client works with a distinct Client-Manager. For example, to enable parallelism in a restricted python environment, it is necessary to use MPI-supported clients coupled with a compatible MPI-based CL-Manager. Similarly, communications with the CL-Manager occur over web protocols when working with IoT clients. Thus, it is imperative to have both components' architecture capable of adapting to diverse circumstances. In Figure \ref{fig:cl-man}, we show the connection between both components and highlight the applicability of a single CL-Manager in supporting multiple types of clients simultaneously.

Finally, the core controls the execution of the given interfaces following the FL workflow discussed in Section \ref{sec:fl-workflow}. It accepts any components extending the FAL protocols with additional subscriber components will discuss in Section \ref{sec:sub}. To keep track of the progress, the FL core creates an FL-Context. It withholds information regarding the state of the execution, including the global model weights, round number, and metrics results after each round's completion.

Regardless of what a researcher aims for, an inclusive environment is vital for running tests and obtaining results while comparing them with others. Through FAL, we provide the necessary environment for researchers to overlook the context and concentrate on their targets. Additionally, the modularity of the framework components, which originated from the layered architecture, allows module swapping, minimizes efforts, and secures a fair comparison.

\subsection{FL-Subscribers} \label{sec:sub}

\begin{figure*}[ht]
\centering
\includegraphics[width=0.9\textwidth]{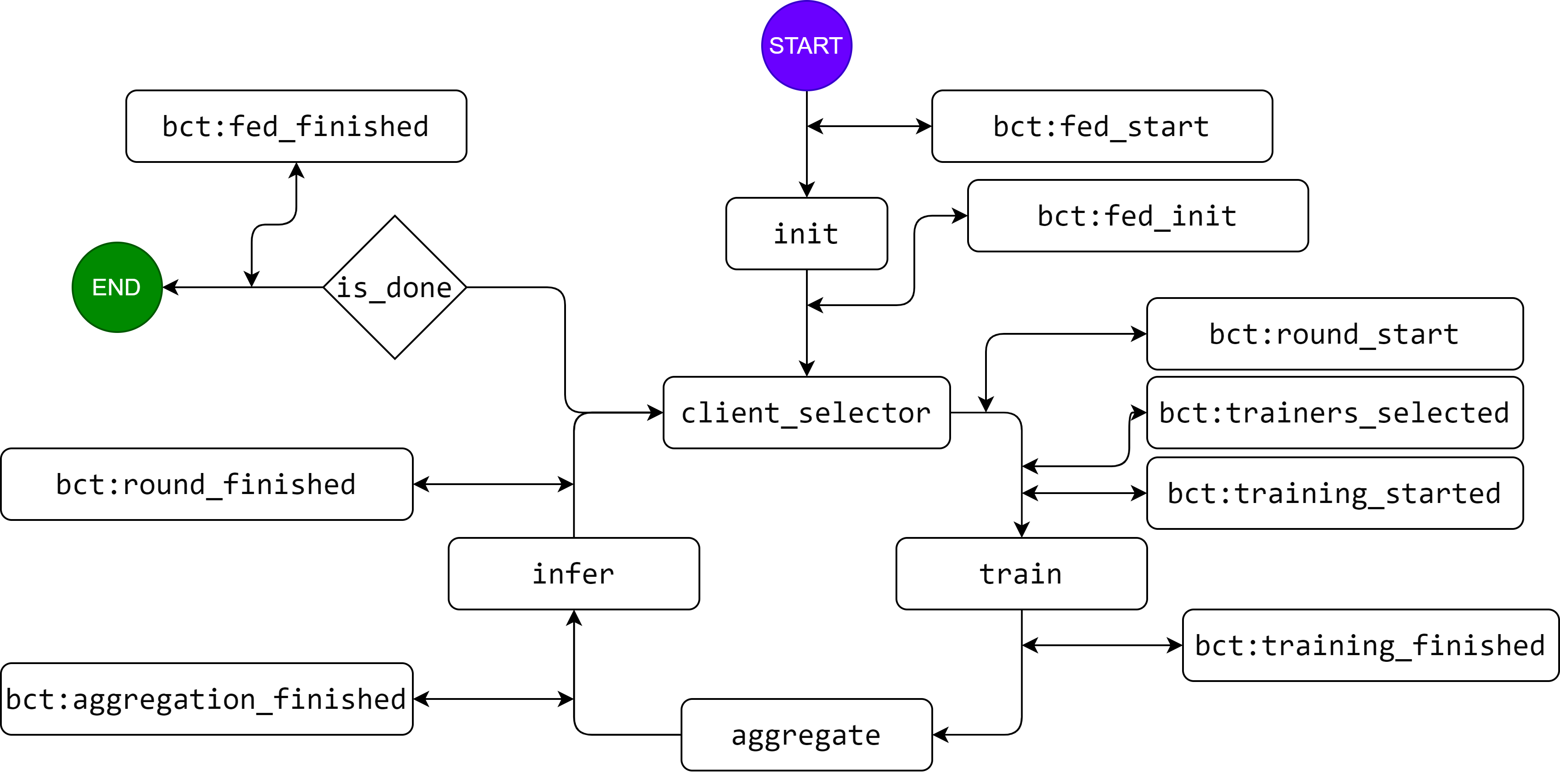}
\caption{Federated Learning Workflow Including Subscribers}
\label{fig:adv-workflow}
\end{figure*}

A typical federated learning application entails different mechanisms such as logging, caching, monitoring the model evolution, drawing charts, saving results, and others. The diversity of these tools is not limited, as each project needs its own. For example, integrating model evolution to monitor the weight divergence between the clients' parameters will eventually increase the workflow intricacy and complicate its integration with other solutions. Thus, we consider our subscription architecture to solve the mentioned challenges. We built our kernel following the observable software design pattern to achieve our objective. During the initialization step, the FL core registers a list of components subject to the FL Subscriber protocols. During the execution, each subscriber receives, in real-time, broadcasts from the core in the form of updates comprising the states of the intended events. These events have predetermined a priory, and each contains an update bound to specific events in the workflow. Figure \ref{fig:adv-workflow} contains a detailed representation of the FL workflow core following the integration of the observable pattern. The events cover every step in the workflow; each provides the execution state and distinct information related to the event phase. For example, $bct:trainers\_selected$ provides information about the selected trainer ids, while $bct:round\_finished$ includes the measured metrics of the Global Model after merging the latest updates.

\begin{figure}[ht]
\centering
\includegraphics[width=0.98\textwidth]{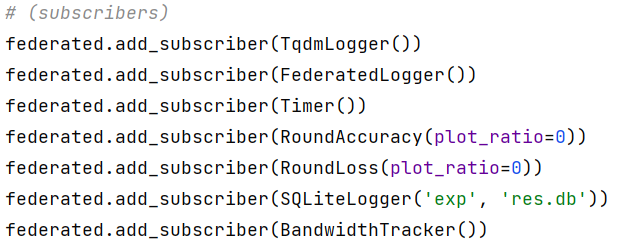}
\caption{Subscription Design Pattern}
\label{fig:code_sub}
\end{figure}

Figure \ref{fig:code_sub} depicts the subscribers integration into a training procedure. The framework provides a list of subscribers, such as Logging Subscribers, to keep track of the execution progress and estimate the execution time. Caching Subscriber is capable of saving execution checkpoints, allowing the runtime to be resumable in case it shuts down due to unplanned circumstances. Metrics Logger stores the metrics result to SQL database, Markup Based Files or external third-party tools such as TensorBoard or Wandb \cite{wandb}. Finally, we include analysis tools as subscribers to monitor the model weights evolution or client selection shift after each federated round. Limitless functionalities can be plugged into our framework without altering the core. It is an excellent choice for researchers aiming for collaboration, extending their work or carrying out their methodologies to other approaches.

\subsection{Data-Center}
Providing solutions for federated learning requires working with different datasets collected from various sources. Additionally, most experiments include working with multiple data distributions between clients to check if the solution is tolerant to the Non-IID problem. Simulating various Non-IID client behaviour is challenging when combined with the diversity of the datasets. Thus, we aim through our Data-Center to provide the necessary protocols addressing the aforementioned problems. Figure \ref{fig:framework} introduces our Data-Center, as an underlying methodology supporting our framework with its three key components: Containers, Distributors, and Providers.

Integrating data \textbf{Containers} unifies the data objects under one protocol across the framework, alleviating the complexity of working with diverse sources. 
Moreover, the core integrates the containers in its workflow, allowing it to handle some dataset routines such as train/test split, batching and type conversion. \textbf{Providers} represent the raw dataset sources used for the client simulation. The source can be anything from cloud storage, databases, local files, and others. The protocols imply outsourcing the raw data to the framework as Containers, which can be used directly to simulate the client's behaviours. Finally, it is essential to demonstrate the capability of any approach against the statistical difficulties caused by the federated learning architecture. With the \textbf{Distributors} component, we aim to standardize the distribution strategies under one protocol capable of simulating most of the Non-IID/IID scenarios. The behaviour of the distributor varies depending on the implementation. In the following, we explain in detail the currently supported distribution while enclosing heatmaps images which present an example of the data distributed to each client following the usage of a distributor. In the images, the X-axis represents the labels, the Y-axis represents the clients, and the intersection between X and Y represents the number of records a label $y$ a client $x$ container withholds. We built a dataset of records identified by $10$ labels distributed to $20$ clients for these representations.

\begin{figure}[h]
\begin{subfigure}{.49\textwidth}
  \centering
  \includegraphics[width=.98\linewidth]{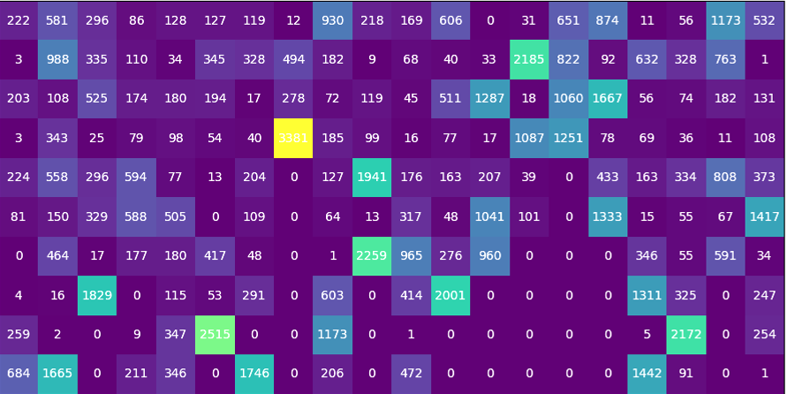}
  \caption{Dirichlet Distribution with $\alpha=0.5$}
  \label{fig:dis:dir05}
\end{subfigure}%
\begin{subfigure}{.49\textwidth}
  \centering
  \includegraphics[width=.98\linewidth]{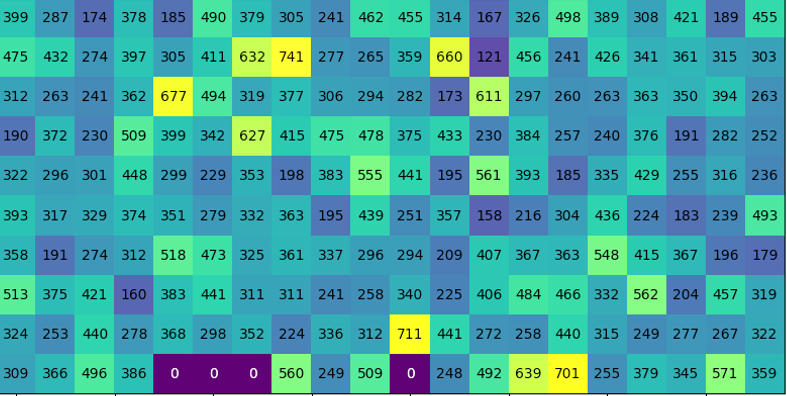}
  \caption{Dirichlet Distribution with $\alpha=10$}
  \label{fig:dis:dir10}
\end{subfigure}
\begin{subfigure}{.49\textwidth}
  \centering
  \includegraphics[width=.98\linewidth]{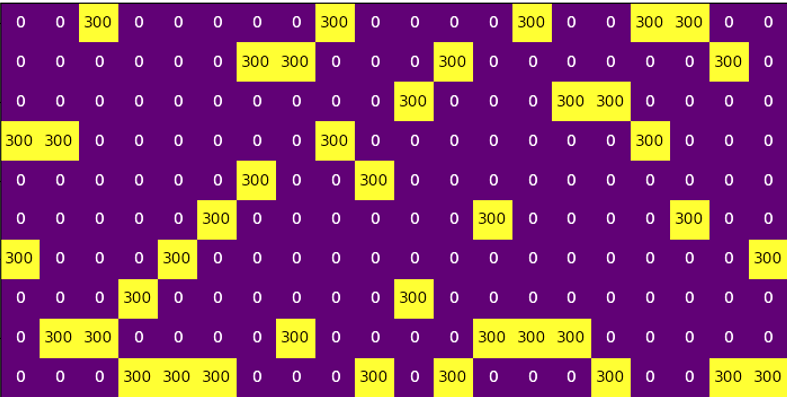}
  \caption{Shard Distribution with $S=2$}
  \label{fig:dis:shard}
\end{subfigure}
\begin{subfigure}{.49\textwidth}
  \centering
  \includegraphics[width=.98\linewidth]{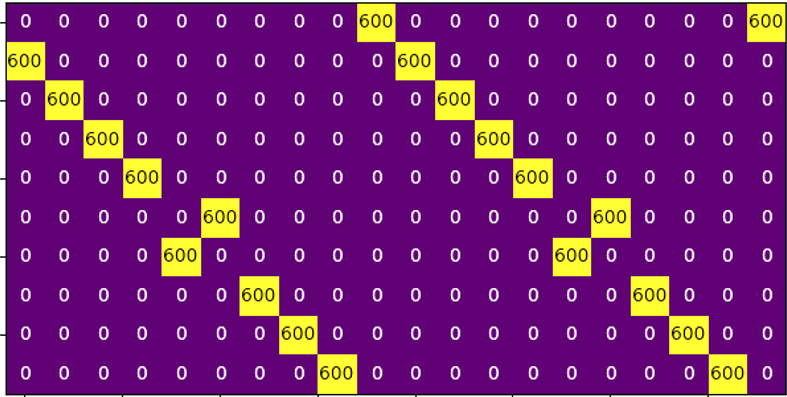}
  \caption{Label Distribution with $L=1$}
  \label{fig:dis:label1}
\end{subfigure}
\begin{subfigure}{.49\textwidth}
  \centering
  \includegraphics[width=.98\linewidth]{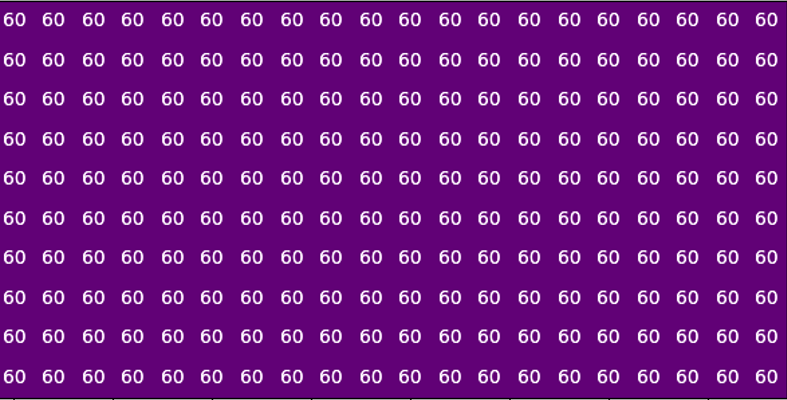}
  \caption{Label Distribution with $L=10$}
  \label{fig:dis:label10}
\end{subfigure}
\begin{subfigure}{.49\textwidth}
  \centering
  \includegraphics[width=.48\linewidth]{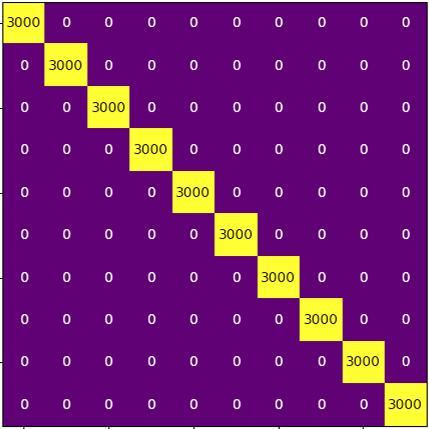}
  \caption{Unique Distribution}
  \label{fig:dis:unique}
\end{subfigure}
\caption{MNIST FL Experiments on Label, Dirichlet, Shard, and Unique Distributions}
\label{fig:fig}
\end{figure}

\textbf{ShardDistributor:}
Introduced by google \cite{google}, such distribution allows experimenting under strict Non-IId environments. The data is split into shards $S$ of equal size. Each shard contains a fixed number of raw records with the same label. Afterward, the shards are distributed to clients, each receiving a predetermined number of shards selected randomly. In \cite{google} use cases, each client receives two shards of 300 records each, indicating that each client has at most two labels in its local datasets, creating a highly Non-IID scenario. It is possible to efficiently control the IIDness severity in this case as it depends on a single value: how many shards $S$ each client receives. The shard distribution is represented in Figure \ref{fig:dis:shard}. Each shard $S$ is fixed with 300 records, and 2 of these shards are distributed to 20 clients.

\textbf{LabelDistributor:} 
Similar to the shard distribution but with more control over the number of labels $L$ distributed to each client. Unlike Shard Distribution, each client is guaranteed to have the specified number of labels in their datasets. Additionally, it's possible to simulate client distribution with different label sizes, which is irrelevant when using Shard Distribution. In Figure \ref{fig:dis:label10} and \ref{fig:dis:label1}, we show the results of a label distribution to 20 clients each receiving 600 records. While Figure \ref{fig:dis:label10} represents an IID distribution where all the clients where all clients receive an equal amount of data for each label, in Figure \ref{fig:dis:label1}, we show a highly Non-IID distribution where each client hold records from only one label.

\textbf{UniqueDistributor:} Using a unique distributor, it is possible to create a particular type of severe Non-IID case in which each client have a dataset with records from a single, unique label/class. Unlike the rest of the distributors, where the same record label/class might exist on multiple client datasets, the Unique distributor guarantees that the single label in one client dataset does not exist in another. The Unique distribution is considered a serious difficulty of Non-IIDness and occurs in studies focusing on clients' penalization, such as behavioural analysis \cite{google}. Such a distribution is presented in Figure \ref{fig:dis:unique}. Different from the highly Non-IID case presented in \ref{fig:dis:label1} where each client holds records from only one label, in the Unique case, two clients holding the same label does not exists. As a result, it is only possible to show the data distribution to 10 clients since our dataset has only ten labels.

\textbf{DirichletDistributor:} applied by \cite{fedml}, used to generate a vector of samples where each signifies the percentage of the representative label index of the total data size, which can be assigned randomly during the distribution. In this case, it is possible to create a form of data unbalancement in terms of local labels across different clients. For instance, we can use Dirichlet to create a client that possesses 10\% of its total records labelled 0, and the rest labelled 1, while another client possesses 85\% of records labelled 0 and the rest labelled 1. Numbers vectors: [0.1,0.9] and [0.85,0.15] are two samples drawn from the Dirichlet Distribution with predefined alpha ($\alpha$) values that denote the data skewness. Such distribution can represent real-life scenarios where data differ not only in terms of general datasets size between clients but also in terms of each label size. In Figure \ref{fig:dis:dir10} we show an IID Dirichlet Distribution with $\alpha=10$ where each client holds almost every label in equal size with little deviation. In Figure \ref{fig:dis:dir05} we show a Non-IID case where $\alpha=0.5$. In this case, a significant number of clients with missing labels combined with a huge divergence in label size across multiple clients.

The combination of the aforementioned mechanics delivers a smooth and straightforward interaction between the datasets and the framework, qualifying the researchers' complete and direct control over the data. For example, unlike other frameworks, the datasets are independent of the core, allowing researchers to take advantage of these benefits when supplementing their datasets. Moreover, the framework supports personalized distributors and providers as long as they integrate the corresponding protocols.

\section{Datasets, Benchmarks \& Experiments}

\subsection{Datasets}
\begin{figure}
\centering
\includegraphics[width=0.78\textwidth]{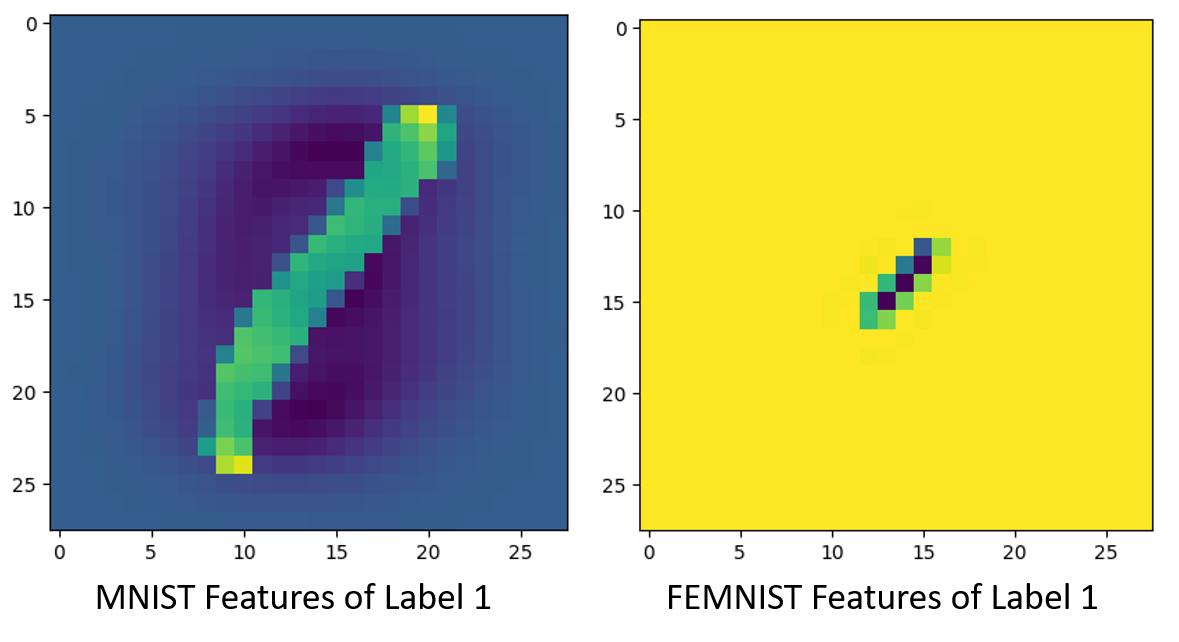}
\caption{Comparison between MNIST and FEMNIST images.}
\label{fig:mnvsfmn}
\end{figure}

To test our framework and confirm its validity, we ran multiple tests, including working with various datasets, models, and configurations. We examine the framework performance on various datasets, including MNIST, FEMNIST, and CIFAR10, covering use cases that could exist in any classification problem. MNIST is a digit dataset which consists of written digit images of numbers from 0 to 9 in two channels. The dataset contains 70k records of 24*24 pixels. On the other hand, FEMNIST is another similar dataset to MNIST but with increased difficulty and more labels, including 28*28 pixel images of letters and digits, forming a dataset with 62 labels distributed between 671k records. The main difference between MNIST and FEMNIST is in the active pixel coverage in both dataset images. Figure \ref{fig:mnvsfmn} shows a comparison between the same digit image. Having fewer active pixels in FEMNIST makes that dataset more challenging and requires a more capable model, which requires additional time and resources. CIFAR10 is a categorical classification dataset containing images of animals and objects in three channels. The main objective is to create a model capable of differentiating between them. The dataset contains 60k 32*32*32 images distributed into $10$ labels. 


Regarding the global model configuration, we used Logistic-Regression for MNIST, a simple and fast model with fewer parameters than others with convoluted configurations. As a result, low-end device trains model faster due to the reduced allocated resources and less bandwidth consumption at the cost of not reaching higher accuracy. It is always the decision of precision vs performance in federated learning where we sacrifice precision for a faster and lightweight model or the contrary. CNN is used for FEMNIST and CIFAR10, each with a different configuration adapting to the differences in the image size. 


\subsection{Configuration Parameters}
Regarding the framework configuration, it consists of the following parameters: 
\begin{itemize}
    \item Data Distributor: It defines how the data is distributed to the simulated clients. This parameter dramatically impacts the model accuracy, such as if the dataset is distributed in an IID or Non-IID manner. The latter's impact is determined by its severity which is discussed in the previous sections. 
    \item CL-Manager: Affect the framework speed and portray the communication between the FL server with its clients. There are two CL-Manager used in the experiments, SequentialManager, which simulates synchronous trainers running in sequence one after another, and an MPIManager, which simulates parallel trainers running concurrently. 
    \item Epoch: Have a significant impact on the framework accuracy. An important term is introduced by \cite{google}, which calls a single epoch and no-batching federated learning FedSGD in which the trainers train their model running over all the data only once and send the model back to the server. In our experiments, we tested the model on both 1 Epoch (FedSGD) and 50 Epoch to show the impact of the parameters on the model accuracy evolution. 
    \item Client Selection: Throughout our experiment, we used a simple client selection procedure: selecting a predefined random number of clients for each round. This parameter simulates real-case scenarios where most clients are not simultaneously available. 
    \item Learning Rate: A hyper-parameter set for model training to control the rate of the model's parameters update and how much the model accepts from the new weights through a value ranging between $0^+ and 1$. Choosing the optimal learning rate can be achieved through small tests (a few rounds) to test the model evolution under various learning rate values. We did these tests beforehand for our experiments and picked the optimal results. 
\end{itemize}

\subsection{Benchmarks \& Experimental Results}

\begin{figure*}
\begin{subfigure}{.49\textwidth}
  \centering
  \includegraphics[width=.98\linewidth]{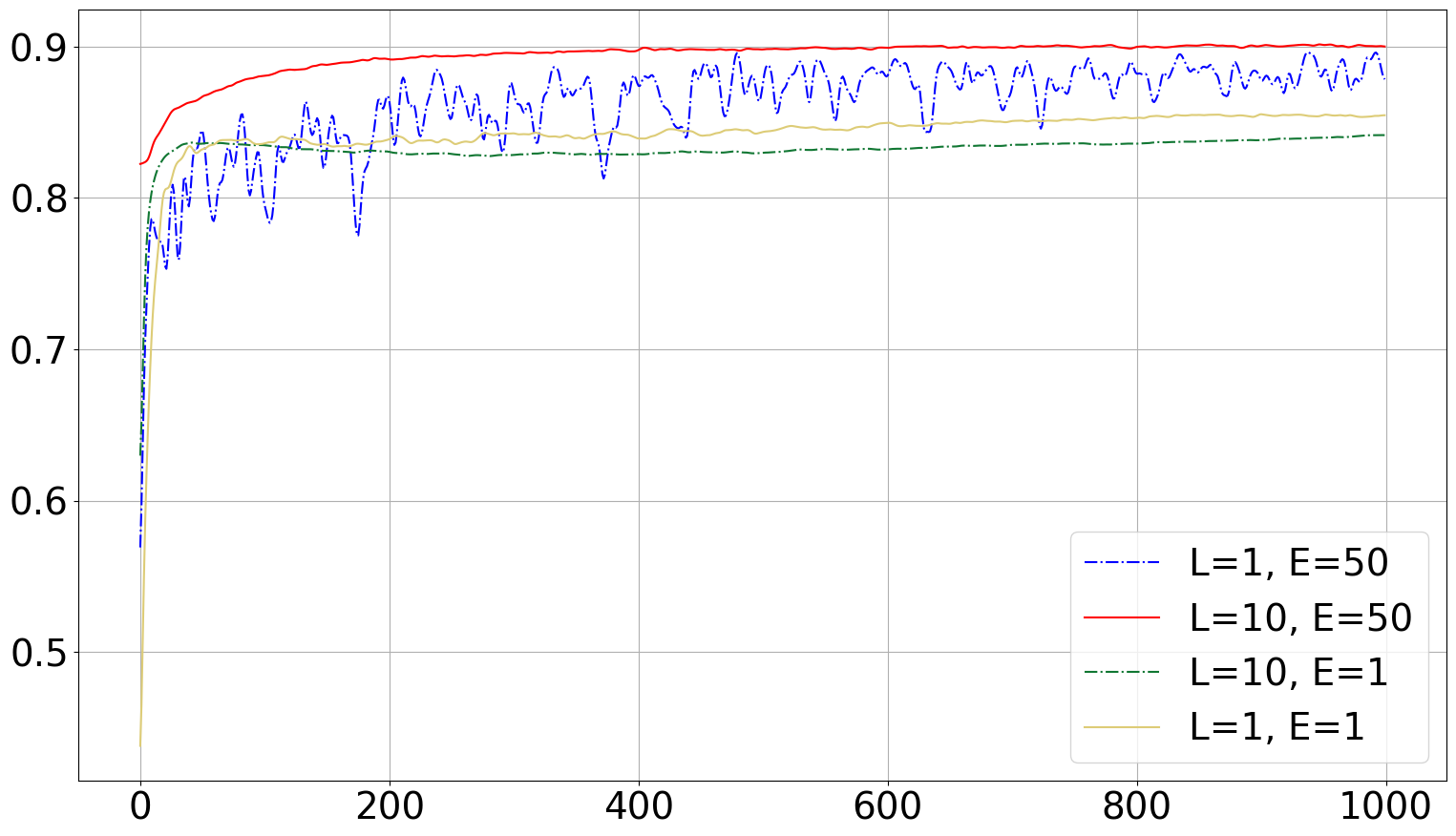}
  \caption{Label Distribution on MNIST Dataset}
  \label{fig:mnist:lbl}
\end{subfigure}%
\begin{subfigure}{.49\textwidth}
  \centering
  \includegraphics[width=.98\linewidth]{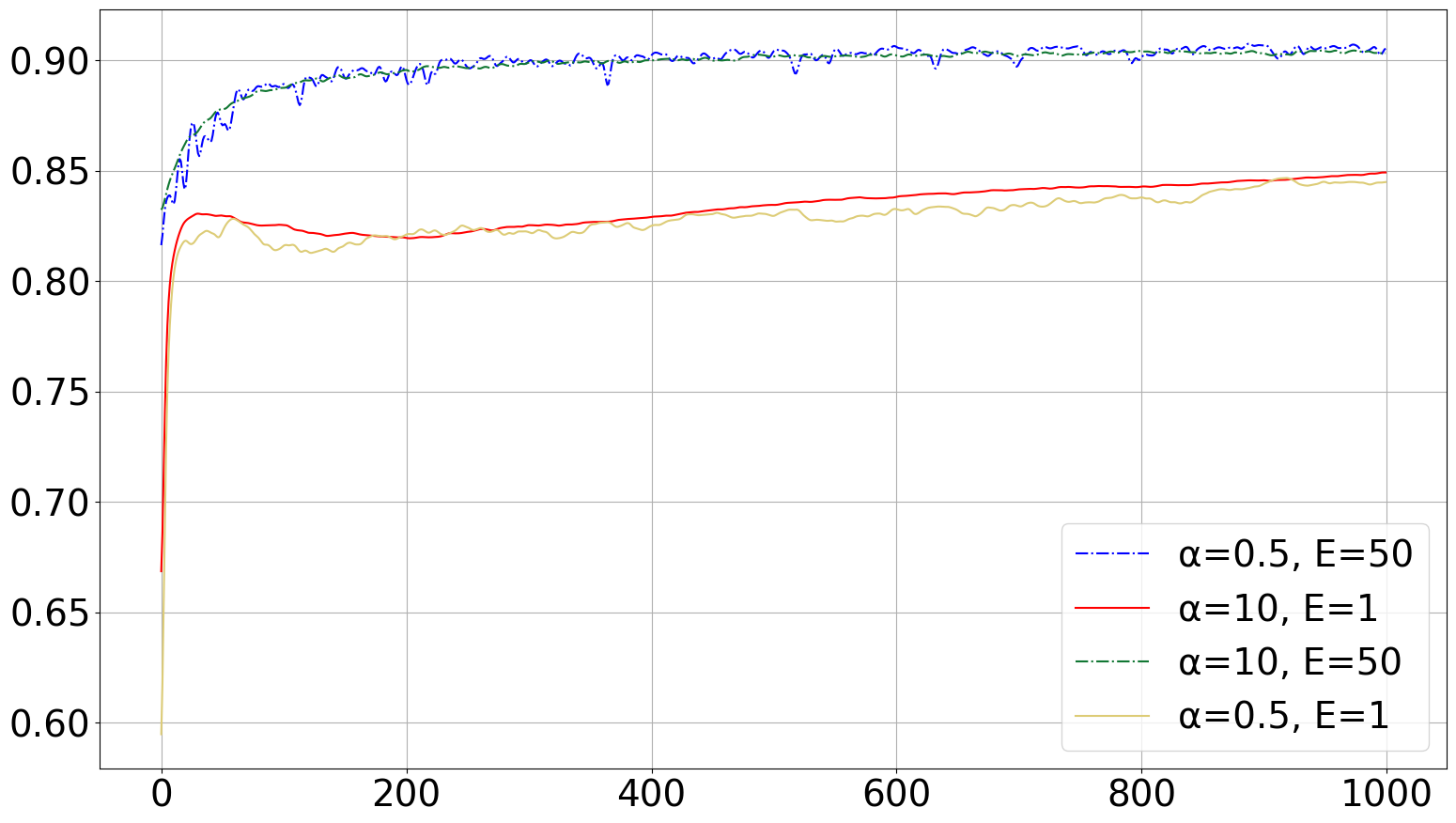}
  \caption{Dirichlet Distribution on MNIST Dataset}
  \label{fig:mnist:dir}
\end{subfigure}
\begin{subfigure}{.49\textwidth}
  \centering
  \includegraphics[width=.98\linewidth]{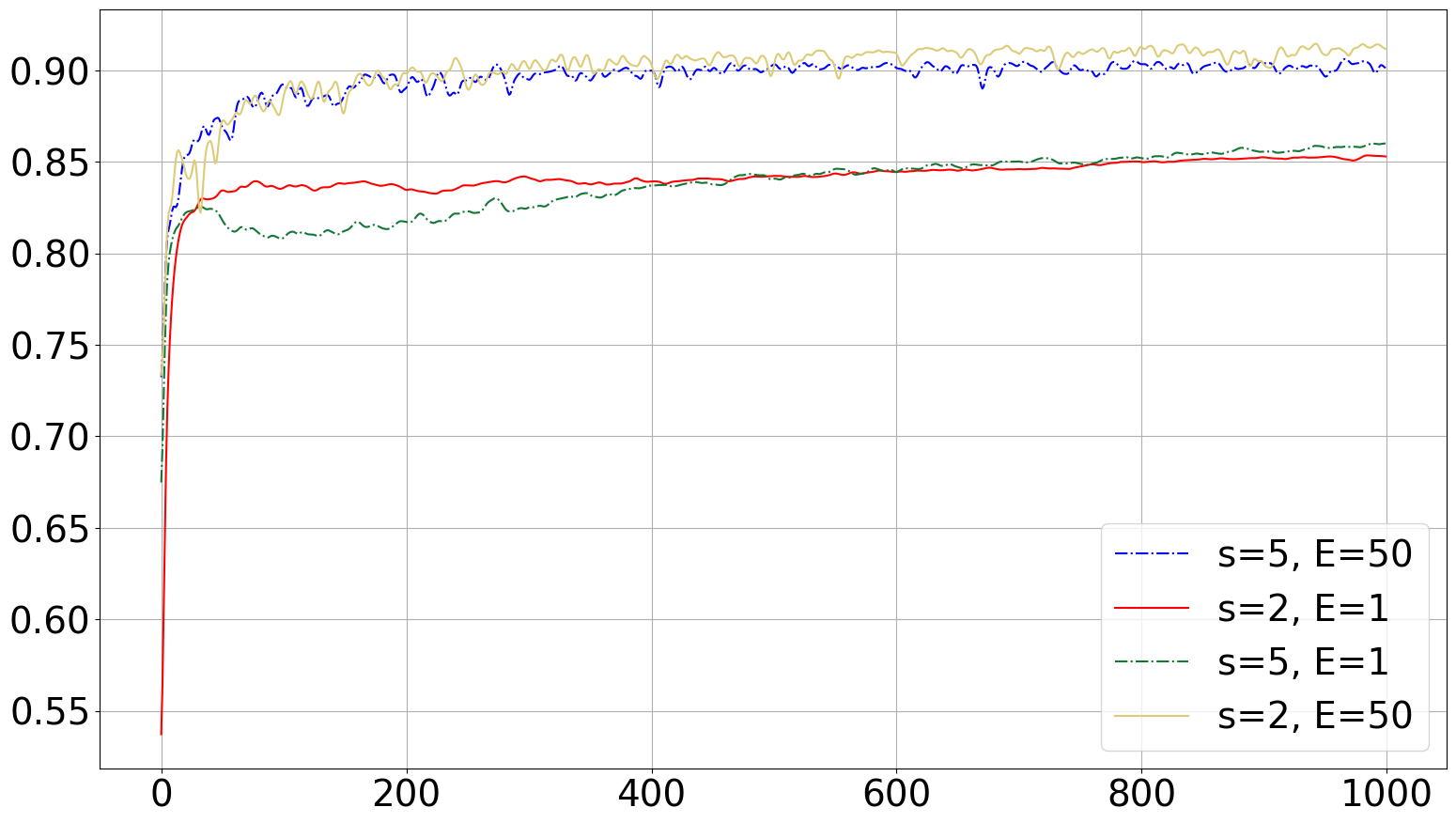}
  \caption{Shard Distribution on MNIST Dataset}
  \label{fig:mnist:shard}
\end{subfigure}
\begin{subfigure}{.49\textwidth}
  \centering
  \includegraphics[width=.98\linewidth]{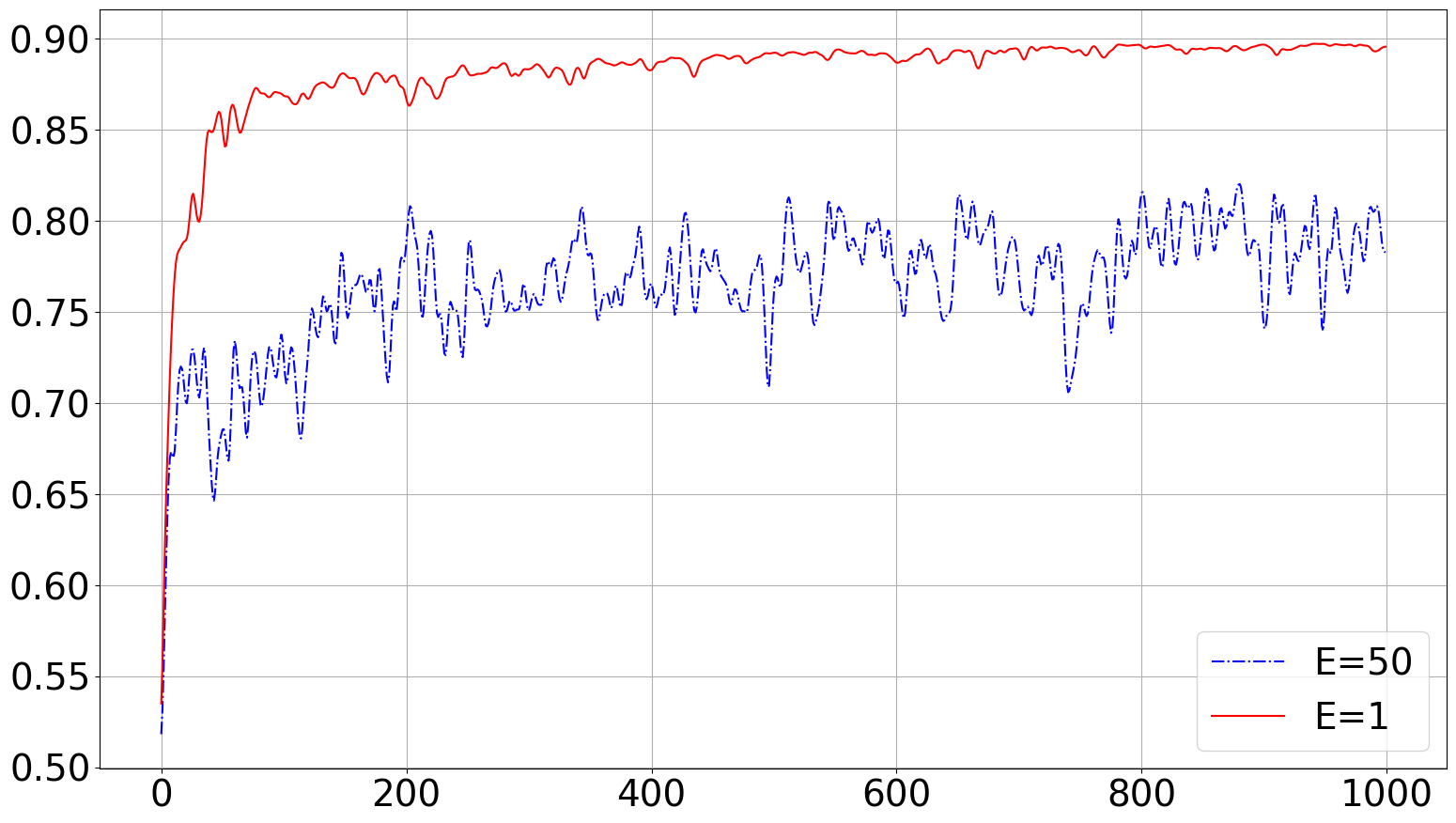}
  \caption{Unique Distribution on MNIST Dataset}
  \label{fig:mnist:unique}
\end{subfigure}
\caption{MNIST FL Experiments on Label, Dirichlet, Shard, and Unique Distributions}
\label{fig:fig_exps_first}
\end{figure*}

Our primary objective is to provide a platform capable of incorporating the huge configurations varieties in the FL environment. Additionally, we aim to effortlessly replicate the main issues in the FL context allowing researchers to solve them in a straightforward and organized manner. In the following experiments, we test and validate our FL framework under various data distributions, which are the roots of the statistical issues in FL, and client-server approaches.

\begin{figure*}
\begin{subfigure}{.49\textwidth}
  \centering
  \includegraphics[width=.98\linewidth]{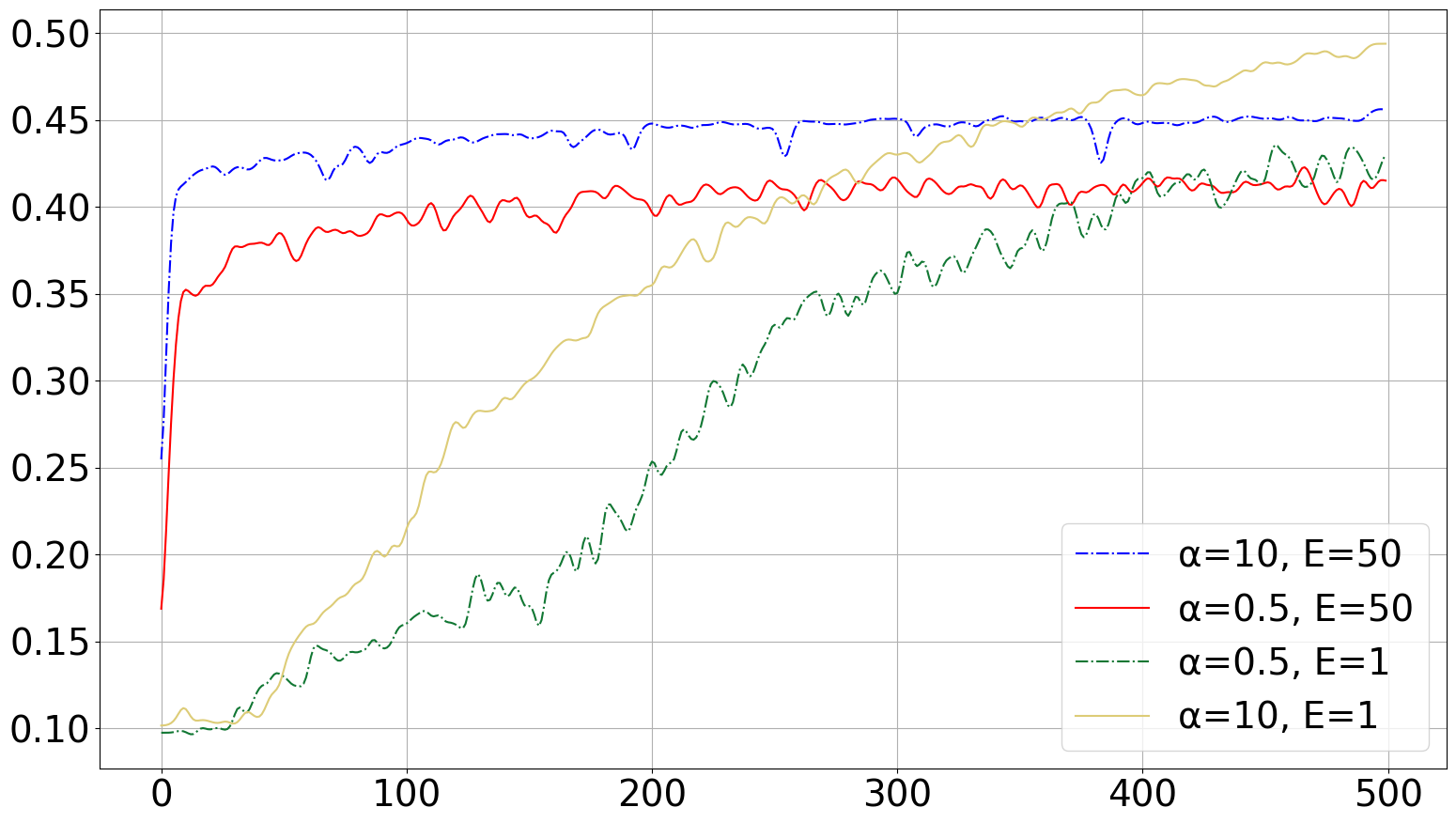}
  \caption{Dirichlet Distribution on CIFAR10 Dataset}
  \label{fig:cifar10:dir}
\end{subfigure}
\begin{subfigure}{.49\textwidth}
  \centering
  \includegraphics[width=.98\linewidth]{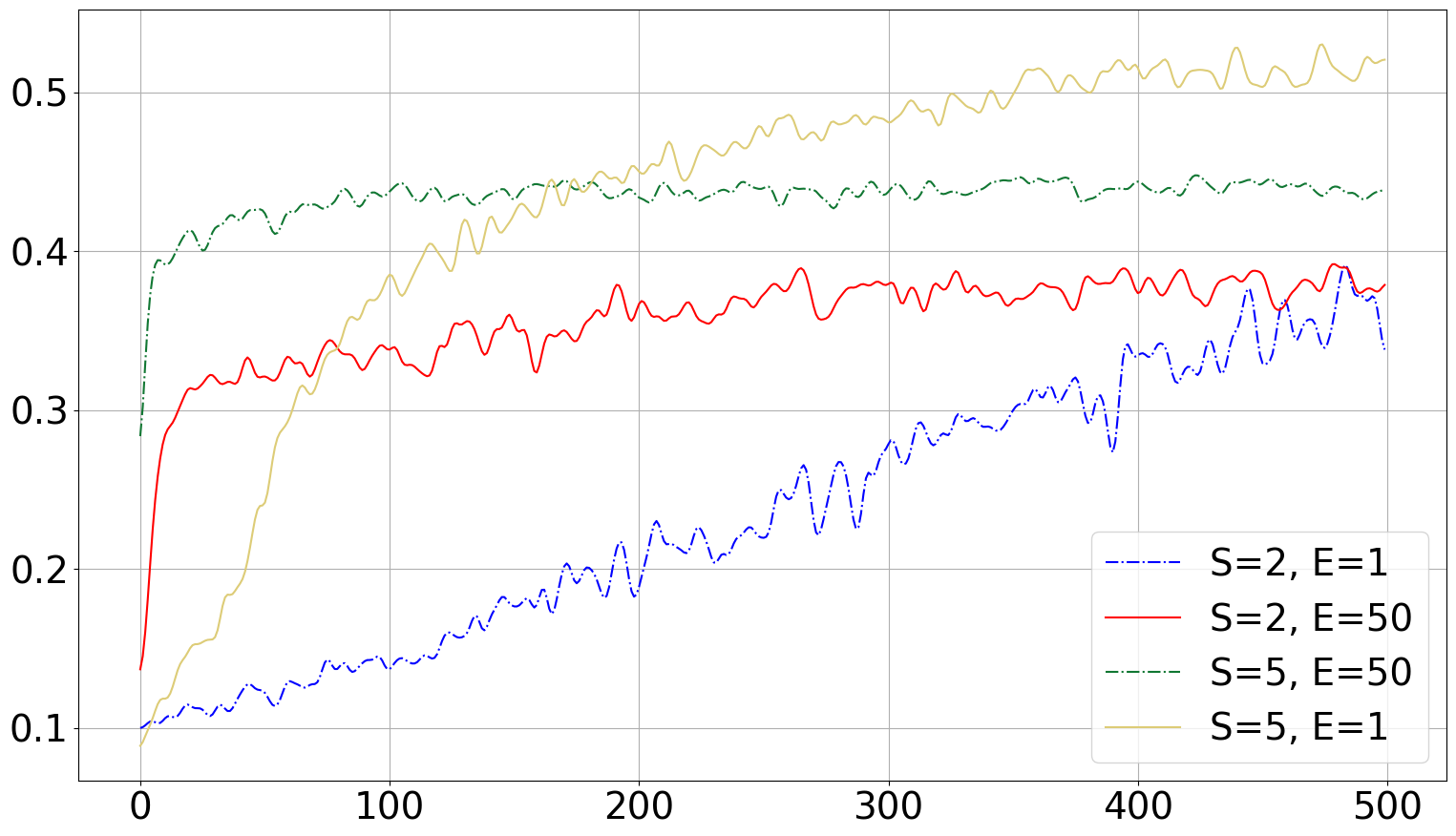}
  \caption{Shards Distribution on CIFAR10 Dataset}
  \label{fig:cifar10:shards}
\end{subfigure}
\begin{subfigure}{.49\textwidth}
  \centering
  \includegraphics[width=.98\linewidth]{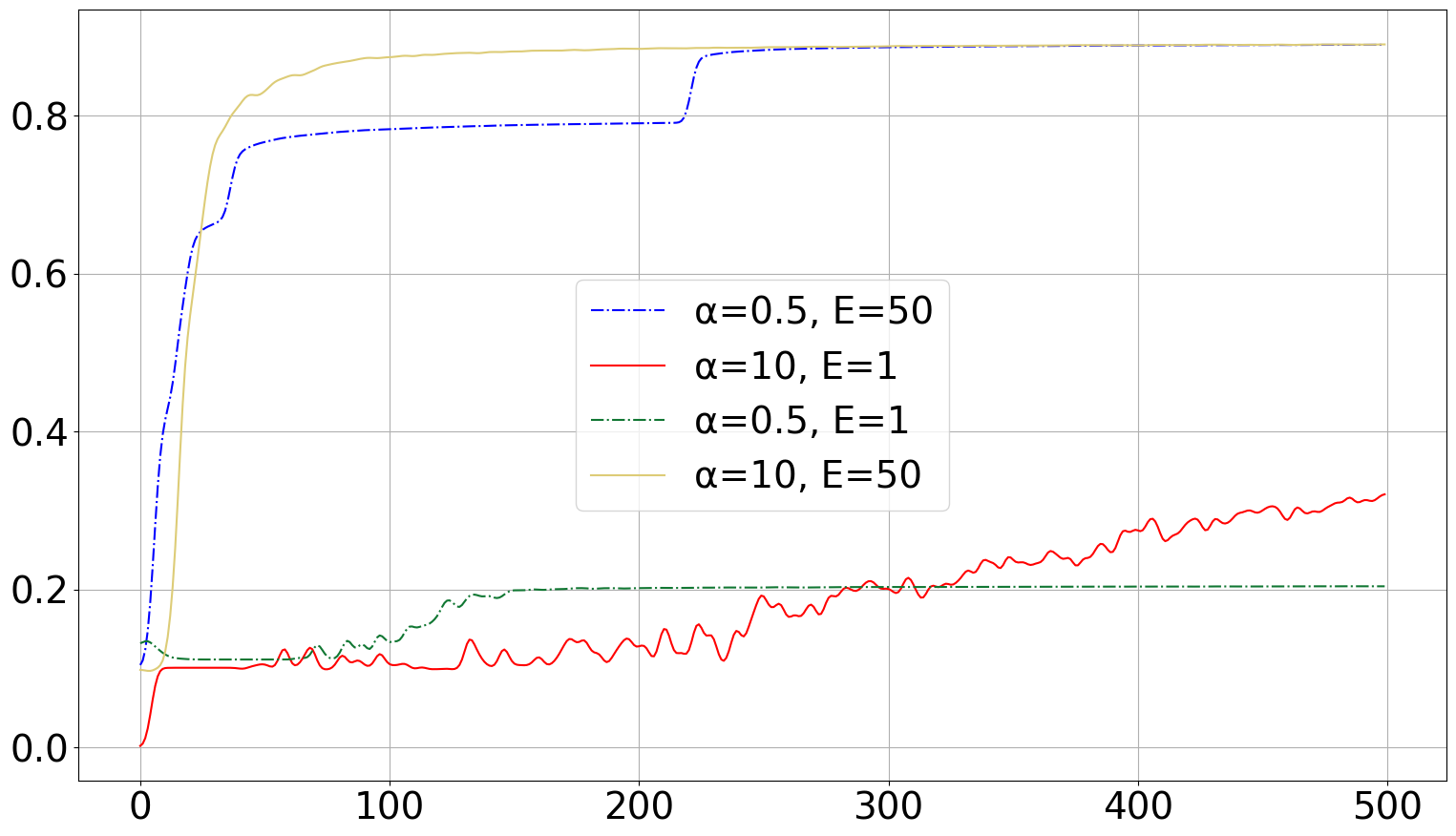}
  \caption{Dirichlet Distribution on FEMNIST Dataset}
  \label{fig:femnist:dir}
\end{subfigure}
\begin{subfigure}{.49\textwidth}
  \centering
  \includegraphics[width=.98\linewidth]{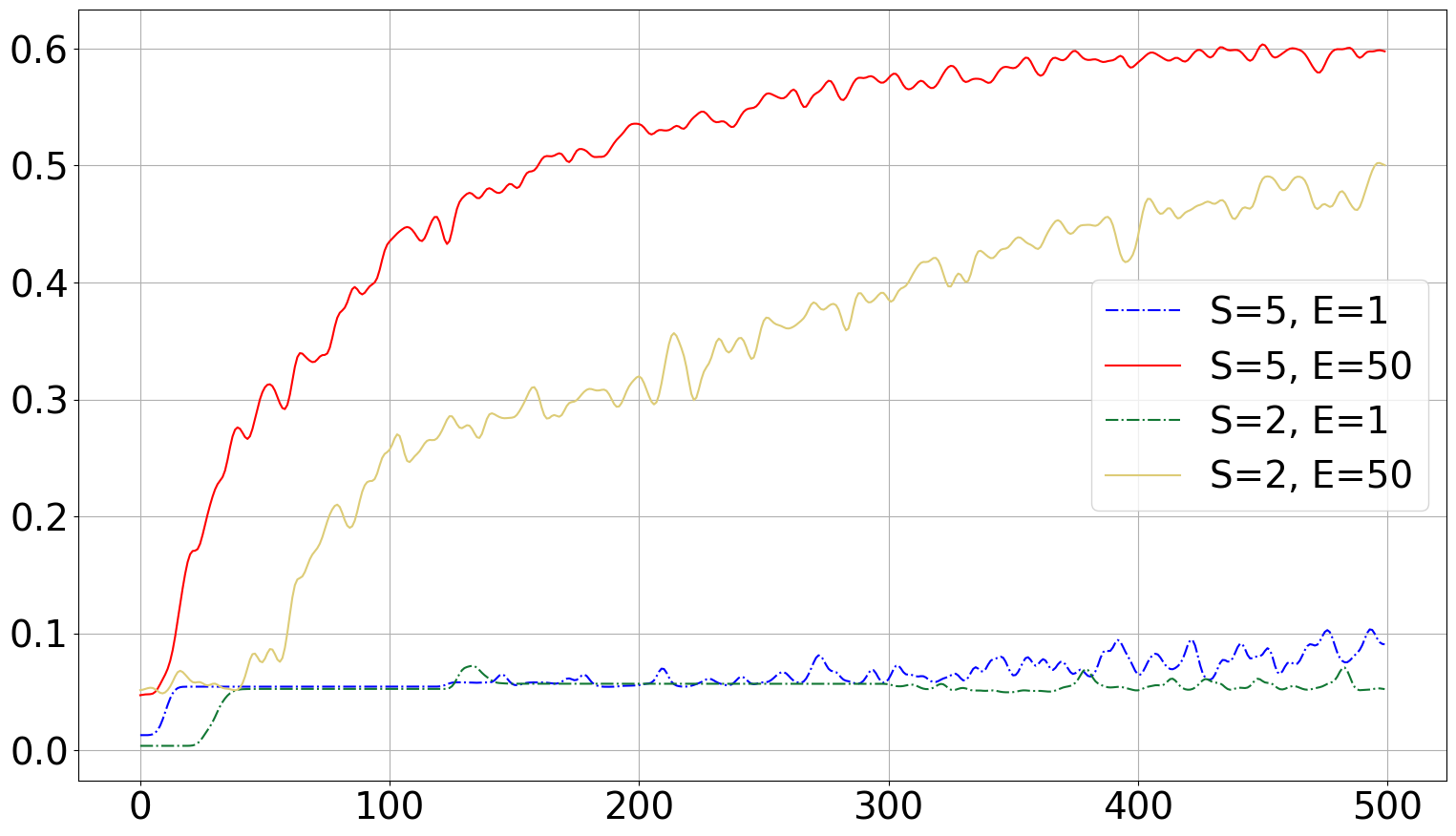}
  \caption{Shards Distribution on FEMNIST Dataset}
  \label{fig:femnist:shards}
\end{subfigure}
\caption{CIFAR and FEMNIST FL Experiments on Dirichlet and Shard Distributions}
\label{fig:fig_exps_second}
\end{figure*}

Figures in \ref{fig:fig_exps_first} and \ref{fig:fig_exps_second} present the evolution of the global model accuracy when experimenting with the impact of both data distribution and epochs. Every test includes FedSGD (single epoch/round) and FedAVG (multiple epochs/round) to show the epoch's influence. Regarding the distribution, we experiment on MNIST using every mentioned distribution to demonstrate and highlight its influence on accuracy. Respecting the limited article length, for the subsequent datasets, we show only Dirichlet due to its capabilities of simulating real-life scenarios and Shard Distribution which is mainly used in FL works. Furthermore, each experiment includes both IID and Non-IID distributions. For Label Distribution, $L=10$ is considered IID, where each client receives data of equal size from every Label, while the $L=1$ is considered the Non-IID case, where each client receives only one Label. For Dirichlet Distribution, we act on the $\alpha$ value, which controls the data skewness of the randomly generated float values. A higher $\alpha$ value results in a notable data disparity between clients, which reduces the data Non-IIDness. Thus, we chose $\alpha=10$ as an IID representative and $\alpha=0.5$ to portray a Non-IID distribution. Finally, for Shard Distribution experiments, we consider $S=2$ as Non-IID in which each client has at most two labels of 300 records each and $S=5$ an IID case where each client has at least five labels of 300 records each. The total number of clients is different depending on the subsequent distribution. For instance, working with Label and Dirichlet Distributions, we can predefine a fixed number of clients, of which we chose 100. Shard distribution depends on the number of records and the shard size. For instance, MNIST, with 60k records divided into 300 records/shards results in 200 shards. Thus, we have 100 clients when $S=2$ compared to 40 when $S=5$. As for Unique Distribution, we are limited by the total number of Labels. For example, in MNIST, since the dataset contains only 10 Labels, we can have at most ten clients. For our experiments, we employed a random client selector in which we selected $CR=10$ clients for each round.

Figure \ref{fig:mnist:lbl} shows the results of MNIST Label Distribution when using Logistic-Regression as the base model configuration. When working with the MNIST dataset, the impact of the distribution is negligible. It can be solved with additional rounds because the dataset is structured with simple and properly filtered images. However, there are still some noticeable differences. With $E=50:L=1$, we notice huge spikes in the accuracies, which indicate that the model is struggling to find the optimal parameters. The intensity of spikes decreased with the subsequent rounds but did not converge even after 1000 rounds. Such a scenario led to additional computational and resource consumption compared to $E=50:L=10$, which converged at 90\% after 200 rounds. Nevertheless, the impact of epochs is more noticeable in the next experiment. For $E=1$, the accuracy were lower than $E=50$. Figure \ref{fig:mnist:dir} shows the results under Dirichlet Distribution while Figure \ref{fig:mnist:shard} show them under Shard Distribution. it is hard to notice major differences as both follow the same pattern. All of the mentioned results show that a high epoch rate follows an increase in the model performance. However, a higher epoch does not strictly reflect better accuracy in every case. In Figure \ref{fig:mnist:unique}, we show a severe case of Non-IIDness in which each client has a unique single label in their dataset. In this special case, FedSGD performed better than FedAVG as well as less spike, which indicates a stable improvement. However, this result is only achieved when using LogisticRegression as a model configuration on the MNIST dataset, which might vary when using other configurations.

In the next batch, we present our experiments using FEMNIST and CIFAR10 datasets. Unlike MNIST, the datasets under consideration contain considerably more information regarding the number of pixels and their variety. For FEMNIST, the image is smaller, while in CIFAR10, the images have three channels instead of 2 in FEMNIST and MNIST. Figure \ref{fig:cifar10:dir} and \ref{fig:cifar10:shards} show the influence of Dirchilet and Shard distribution on the accuracy of the federated learning global model using the CIFAR10 dataset. The number of Epochs has a significant impact on these experiments. Under FedSGD, the experiments under both distributions started with lower accuracy and reached a higher accuracy than the $E=50$ experiments.
Additionally, both experiments' accuracies suffered from the Non-IID distribution. Regarding FEMNSIT dataset experiments, Figure \ref{fig:femnist:shards} shows the data distribution of FEMNIST when using the Shard Distributor, while Figure \ref{fig:femnist:dir} portrays the Dirichlet distributor. Under the FedSGD environment, FEMNIST suffers from an apparent problem: the accuracy did not improve in either the IID or Non-IID context. However, increasing the number of Epochs sustains the accuracy improvements. Under Dirichlet influence, the Non-IID distribution started with a worse accuracy while reaching similar results at the end of the experiments. The difference between IID and Non-IID is evident in the shard distribution, Figure \ref{fig:femnist:shards}. 

We compile all of our experimental results in Table \ref{tab:exp:results} showing the final accuracy and loss at the end of each experiment when using $E=1$ and $E=50$ under both IID and Non-IID distributions.

\begin{figure}[h]
\centering
\includegraphics[width=0.78\textwidth]{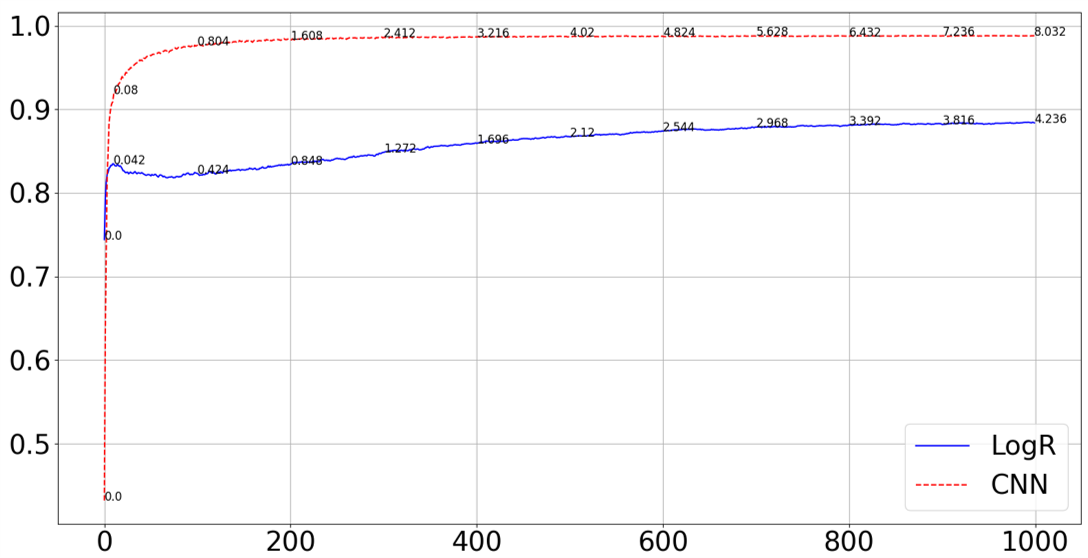}
\caption{Bandwidth transmission cost comparison between CNN and LogisticRegression (LogR) model configuration}
\label{fig:band}
\end{figure}

\begin{table}[H]
\caption{Last Accuracy and Loss of federated learning experiments when using $E=1$ and $E=50$ under IID and Non-IID experiments}
\label{tab:exp:results}
\def\arraystretch{1.5}%
\begin{tabular}{ccl|cccc|}
\cline{4-7}
\multirow{3}{*}{\textbf{}}                              & \multirow{3}{*}{\textbf{}}                      & \multicolumn{1}{c|}{\multirow{3}{*}{\textbf{}}} & \multicolumn{4}{c|}{\textbf{Results}}                                                                                      \\ \cline{4-7} 
                                                        &                                                 & \multicolumn{1}{c|}{}                           & \multicolumn{2}{c|}{\textbf{ACC}}                                      & \multicolumn{2}{c|}{\textbf{LOSS}}                \\ \cline{4-7} 
                                                        &                                                 & \multicolumn{1}{c|}{}                           & \multicolumn{1}{c|}{\textbf{E=1}} & \multicolumn{1}{c|}{\textbf{E=50}} & \multicolumn{1}{c|}{\textbf{E=1}} & \textbf{E=50} \\ \hline
\multicolumn{1}{|c|}{\multirow{7}{*}{\textbf{MNIST}}}   & \multicolumn{1}{c|}{\multirow{2}{*}{Label}}     & IID                                             & \multicolumn{1}{c|}{0.8415}       & \multicolumn{1}{c|}{0.899}         & \multicolumn{1}{c|}{1.621}        & 1.556         \\ \cline{3-7} 
\multicolumn{1}{|c|}{}                                  & \multicolumn{1}{c|}{}                           & N-IID                                           & \multicolumn{1}{c|}{0.854}        & \multicolumn{1}{c|}{0.867}         & \multicolumn{1}{c|}{1.62}         & 1.637         \\ \cline{2-7} 
\multicolumn{1}{|c|}{}                                  & \multicolumn{1}{c|}{\multirow{2}{*}{Shard}}     & IID                                             & \multicolumn{1}{c|}{0.859}        & \multicolumn{1}{c|}{0.9002}        & \multicolumn{1}{c|}{1.593}        & 1.555         \\ \cline{3-7} 
\multicolumn{1}{|c|}{}                                  & \multicolumn{1}{c|}{}                           & N-IID                                           & \multicolumn{1}{c|}{0.852}        & \multicolumn{1}{c|}{0.911}         & \multicolumn{1}{c|}{1.6214}       & 1.573         \\ \cline{2-7} 
\multicolumn{1}{|c|}{}                                  & \multicolumn{1}{c|}{\multirow{2}{*}{Dirichlet}} & IID                                             & \multicolumn{1}{c|}{0.849}        & \multicolumn{1}{c|}{0.903}         & \multicolumn{1}{c|}{1.612}        & 1.554         \\ \cline{3-7} 
\multicolumn{1}{|c|}{}                                  & \multicolumn{1}{c|}{}                           & N-IID                                           & \multicolumn{1}{c|}{0.845}        & \multicolumn{1}{c|}{0.906}         & \multicolumn{1}{c|}{1.613}        & 1.56          \\ \cline{2-7} 
\multicolumn{1}{|c|}{}                                  & \multicolumn{1}{c|}{Unique}                     & N-IID                                          & \multicolumn{1}{c|}{0.896}        & \multicolumn{1}{c|}{0.784}         & \multicolumn{1}{c|}{1.582}        & 1.727         \\ \hline
\multicolumn{1}{|c|}{\multirow{4}{*}{\textbf{CIFAR}}}   & \multicolumn{1}{c|}{\multirow{2}{*}{Shard}}     & IID                                             & \multicolumn{1}{c|}{0.519}        & \multicolumn{1}{c|}{0.438}         & \multicolumn{1}{c|}{1.419}        & 5.104         \\ \cline{3-7} 
\multicolumn{1}{|c|}{}                                  & \multicolumn{1}{c|}{}                           & N-IID                                           & \multicolumn{1}{c|}{0.323}        & \multicolumn{1}{c|}{0.394}         & \multicolumn{1}{c|}{1.996}        & 3.175         \\ \cline{2-7} 
\multicolumn{1}{|c|}{}                                  & \multicolumn{1}{c|}{\multirow{2}{*}{Dirichlet}} & IID                                             & \multicolumn{1}{c|}{0.4946}       & \multicolumn{1}{c|}{0.456}         & \multicolumn{1}{c|}{1.4}          & 5.769         \\ \cline{3-7} 
\multicolumn{1}{|c|}{}                                  & \multicolumn{1}{c|}{}                           & N-IID                                           & \multicolumn{1}{c|}{0.433}        & \multicolumn{1}{c|}{0.415}         & \multicolumn{1}{c|}{1.584}        & 4.719         \\ \hline
\multicolumn{1}{|c|}{\multirow{4}{*}{\textbf{FEMNIST}}} & \multicolumn{1}{c|}{\multirow{2}{*}{Shard}}     & IID                                             & \multicolumn{1}{c|}{0.106}        & \multicolumn{1}{c|}{0.585}         & \multicolumn{1}{c|}{4.068}        & 3.579         \\ \cline{3-7} 
\multicolumn{1}{|c|}{}                                  & \multicolumn{1}{c|}{}                           & N-IID                                           & \multicolumn{1}{c|}{0.0509}       & \multicolumn{1}{c|}{0.498}         & \multicolumn{1}{c|}{4.094}        & 3.691         \\ \cline{2-7} 
\multicolumn{1}{|c|}{}                                  & \multicolumn{1}{c|}{\multirow{2}{*}{Dirichlet}} & IID                                             & \multicolumn{1}{c|}{0.326}        & \multicolumn{1}{c|}{0.889}         & \multicolumn{1}{c|}{3.858}        & 3.265         \\ \cline{3-7} 
\multicolumn{1}{|c|}{}                                  & \multicolumn{1}{c|}{}                           & N-IID                                           & \multicolumn{1}{c|}{0.203}        & \multicolumn{1}{c|}{0.889}         & \multicolumn{1}{c|}{3.951}        & 3.26          \\ \hline
\end{tabular}
\end{table}
Different from our previous experiments, In Figure \ref{fig:band} we highlight the bandwidth cost of using different model configurations showing the impact on the model accuracy. The Figure shows the evolution of the accuracy throughout the federated learning rounds, and the accumulative bandwidth cost annotated for every 100 rounds. The model configurations in use are CNN 2 Layers and LogisticRegression. For the hyper-parameters, we used FedSGD on the MNIST dataset distributed to 100 clients using Dirichlet Distributor with $\alpha=10$. Ten clients are randomly selected to train a model using the provided configuration in each round. Since we are not using any compression algorithm, the models' weight size is constant. In this experiment, using LogisticRegression, we achieved an accuracy of 88\% after 1000 rounds which cost a total of 4.236 MB, compared to 90\% accuracy reached when using CNN achieved after ten rounds with a total cost of 0.08 MB. In this case, an exemplary model configuration can reduce the total bandwidth cost, although it costs more per client per round.

\begin{figure}[h]
\centering
\includegraphics[width=0.78\textwidth]{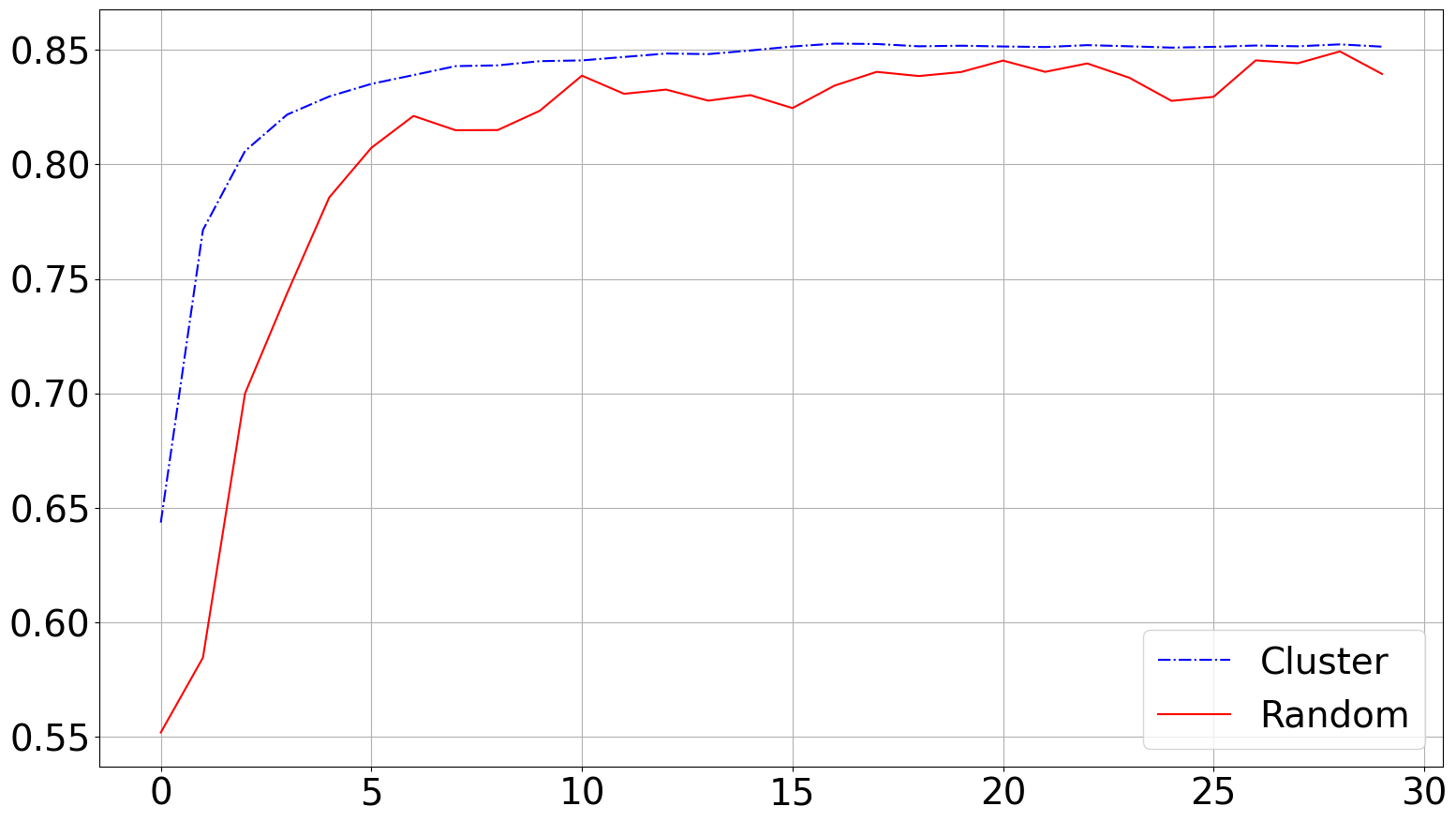}
\caption{Comparison between Cluster Selector and Random Selector using MNIST dataset}
\label{fig:selector}
\end{figure}

In Figure \ref{fig:selector}, we show the impact of the client selector on the global model accuracy after each round. For this experiment, we used Label Distributor with $L=2$ to create a Non-IID data distribution between clients coupled with FedSGD configuration and a client ratio of 10. We compare between random client selector with a cluster-based client selector. In the latter, we start with an initialization round, asking each client to train a model and send it to the server. The selector will use the model as input to identify the client cluster using the K-Means algorithm with a parameter $K$, which refers to the number of clusters. During each round, the clustering algorithm selects a predefined number of clients from each cluster to participate. For instance, for a client ratio of ten, and $K=5$, we select two clients from each cluster. Using such an approach makes it possible to have a variety of model weights for each round, which can be used to solve the Non-IID distribution issues. Overall, the results in Figure \ref{fig:selector} demonstrate the capabilities of the cluster selector to achieve better and more stable results than the random selector.

\subsection{Weight Divergence Analysis Module}

\begin{figure}[h]

\begin{subfigure}{.49\textwidth}
  \centering
  \includegraphics[width=\linewidth]{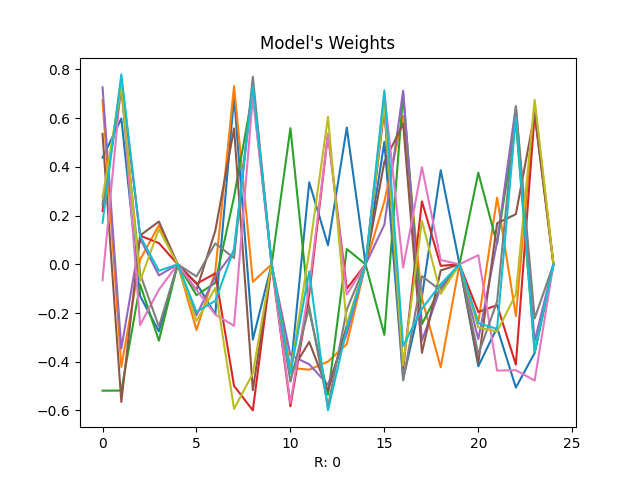}
  \captionsetup{justification=centering}
  \caption{IID \\ Label Distribution $R=0$}
  \label{fig:module:iid1}
\end{subfigure}%
\begin{subfigure}{.49\textwidth}
  \centering
  \includegraphics[width=\linewidth]{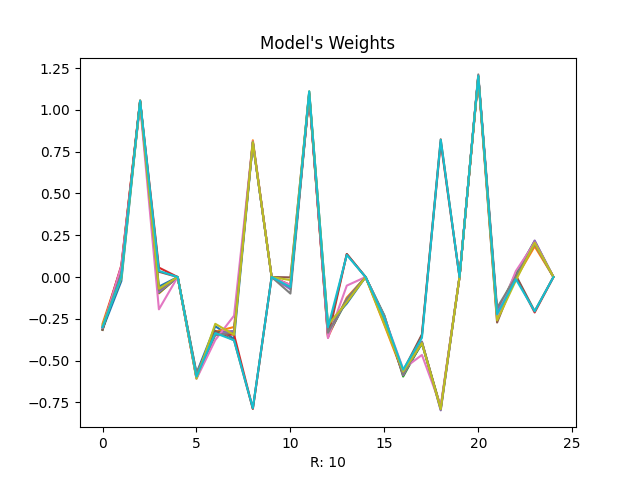}
  \captionsetup{justification=centering}
  \caption{IID \\ Label Distribution $R=10$}
  \label{fig:module:iid2}
\end{subfigure}

\begin{subfigure}{.49\textwidth}
  \centering
  \includegraphics[width=\linewidth]{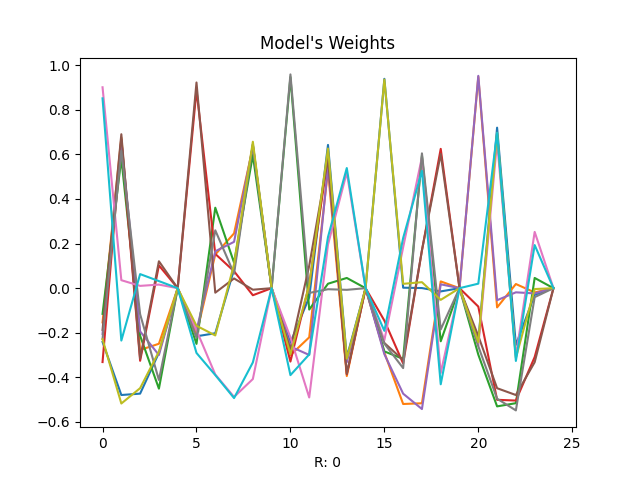}
  \captionsetup{justification=centering}
  \caption{Non-IID \\ Label Distribution $R=0$}
  \label{fig:module:niid1}
\end{subfigure}%
\begin{subfigure}{.49\textwidth}
  \centering
  \includegraphics[width=\linewidth]{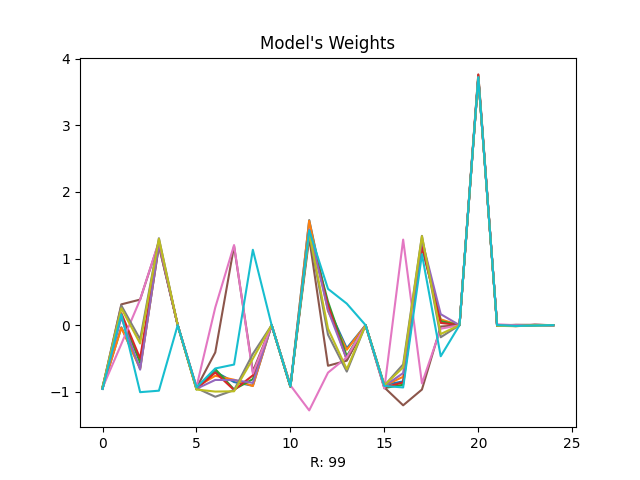}
  \captionsetup{justification=centering}
  \caption{Non-IID \\ Label Distribution $R=99$}
  \label{fig:module:niid2}
\end{subfigure}

\caption{Models' Weights Representation of 100 IID/Non-IID Clients Using Weight Divergence Analysis Module}
\label{fig:module:iid}
\end{figure}

Figures \ref{fig:module:iid} depict the integration of a complex weight analysis mechanism into any federated learning approach. In this experiment, the module shows the evolution of weight divergence between clients' weights in both IID in Figures \ref{fig:module:iid1}, \ref{fig:module:iid2} and Non-IID in Figures \ref{fig:module:niid1}, \ref{fig:module:niid2}. In each figure, each plot line represents the flattened weights of an MNIST client. Due to the size of the weights, principal component analysis (PCA) is used to reduce the number of representable values. 

Comparing the IID experiment \ref{fig:module:iid1} to the Non-IID \ref{fig:module:niid1} at round 0, we can notice a difference in the weight divergence between the clients. When following and IID distribution between clients' data, there tend to exist similarities between the generated models' weights. The appearing weight divergences subside to almost identical weights after $10$ rounds, as shown in Figure \ref{fig:module:iid2}.

On the other hand, the Non-IID clients in Figure \ref{fig:module:niid1} show a higher weight divergence between them. The weight divergence is still evident even after $99$ rounds, as shown in Figure \ref{fig:module:niid2} compared to the IID case. This weight representation can imply a direct connection between the federated learning performance and the weight divergence that exists due to the data distribution. The higher the weight divergence, the more it has adverse effects on the global model in terms of accuracy and convergence rate.

\section{Conclusion}
With federated learning getting acknowledged for its premise, there is a lack of fundamental conventions in available frameworks. We aim to solve this issue by introducing our protocol-based layered architecture, with modular support allowing our framework to act as a complete ecosystem. Using our modular implementation, we consider FL functionalities as separate modules that can be transferable mini-application. Such modularity entitles us to have a junction in our ideas. Additionally, our framework is built from scratch to support projects' extendability while providing the necessary tools to replicate the majority of FL-related issues. We laid the groundwork through ModularFed, and we plan to further enhance it in the future by integrating various supporting technologies as modules. Technologies such as traceability through blockchain or components deployments and orchestration through Kubernetes microservices. Finally, we experimented with the validity and flexibility of our framework under various FL scenarios, including distribution issues, network consumption, and client quality. Our framework will be constantly maintained by introducing new features and various new modules following the latest directions.

\bibliography{main}
\clearpage
\hfill
\section*{Biographies}

\authorbibliography[scale=1, wraplines=4, overhang=40pt, imagewidth=0.25\textwidth, imagepos=r]
{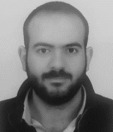}{Mohamad Arafeh}{
Is currently pursuing the Ph.D. degree with École de Technologie Supérieure, Montreal, QC, Canada. He is working in the area of federated learning.\\\\\\\\}

\authorbibliography[scale=0.15, wraplines=8, overhang=40pt, imagewidth=0.25\textwidth, imagepos=r]
{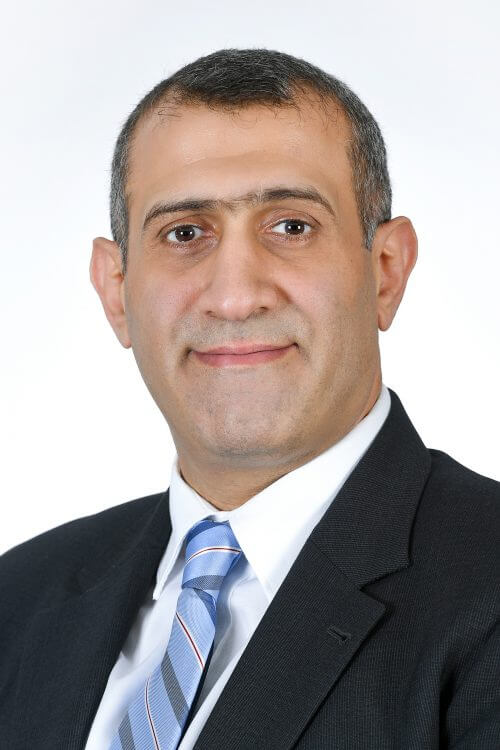}{Hadi Otrok}{
Received the Ph.D. degree in computer science and software engineering from Laval University, Canada, in 2005. He is a Professor with Concordia Institute for Information Systems Engineering, Concordia University, Canada. From 2005 to 2006, he was a Postdoctoral Fellow with Laval University, and then NSERC Postdoctoral Fellow at Simon Fraser University, Canada. He is an NSERC Co-Chair for Discovery Grant for Computer Science (2016-2018). His research interests include the areas of computational logics, model checking, multi-agent systems, services computing, game theory, and deep learning.}

\authorbibliography[scale=0.25, wraplines=8, overhang=40pt, imagewidth=0.25\textwidth, imagepos=r]
{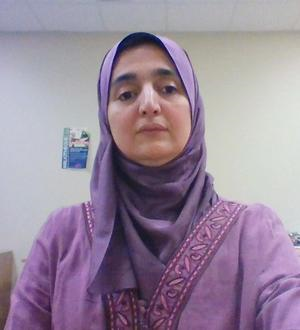}{Hakima Ould-Slimane}{
Obtained her Ph.D. degree in Computer Science from Laval University, Quebec, Canada. She is currently a professor at the department of mathematics and computer science at Universite de Quebec a Trois-Rivieres (UQTR, Trois-Rivieres, Canada). Her research interests include mainly: information security, cyber resilience, homomorphic encryption, federated learning, preserving data privacy in smart environments, machine learning based intrusion detection, access control, optimization of security mechanisms and security of social networks.}

\authorbibliography[scale=0.1, wraplines=8, overhang=40pt, imagewidth=0.25\textwidth, imagepos=r]
{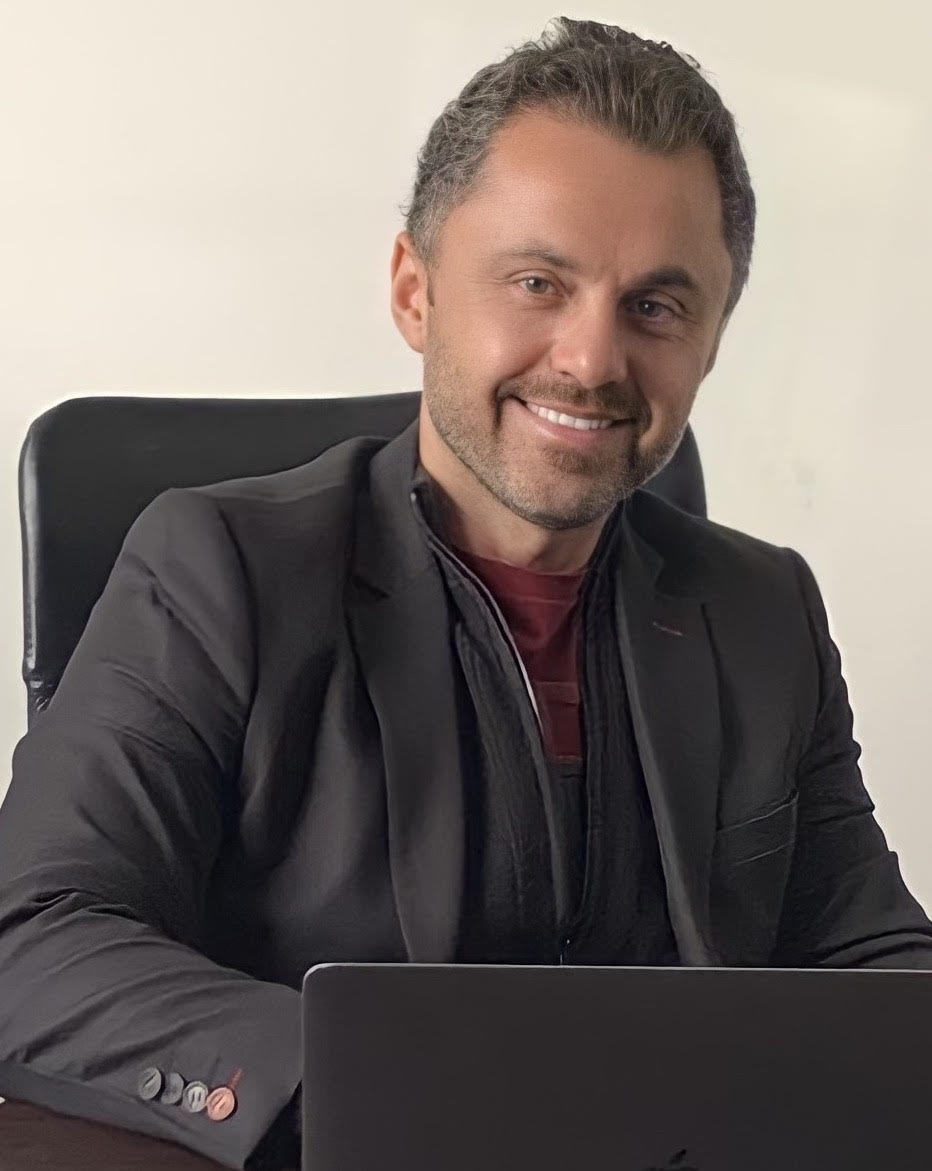}{Azzam Mourad}{
Received his M.Sc. in CS from Laval University, Canada (2003) and Ph.D. in ECE from Concordia University, Canada (2008). He is currently Professor of Computer Science and Founding Director of the Cyber Security Systems and Applied AI Research Center with the Lebanese American University, Visiting Professor of Computer Science with New York University Abu Dhabi and Affiliate Professor with the Software Engineering and IT Department, Ecole de Technologie Superieure (ETS), Montreal, Canada. His research interests include Cyber Security, Federated Machine Learning, Network and Service Optimization and Management targeting IoT and IoV, Cloud/Fog/Edge Computing, and Vehicular and Mobile Networks. He has served/serves as an associate editor for IEEE Transactions on Services Computing, IEEE Transactions on Network and Service Management, IEEE Network, IEEE Open Journal of the Communications Society, IET Quantum Communication, and IEEE Communications Letters, the General Chair of IWCMC2020, the General Co-Chair of WiMob2016, and the Track Chair, a TPC member, and a reviewer for several prestigious journals and conferences. He is an IEEE senior member.}

\authorbibliography[scale=0.3, wraplines=8, overhang=40pt, imagewidth=0.25\textwidth, imagepos=r]
{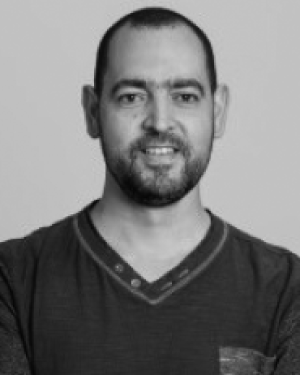}{Chamseddine Talhi}{
Received the Ph.D. degree in computer science from Laval University, Quebec, QC, Canada, in 2007. He is a Professor with the Department of Software Engineering and IT, ÉTS, University of Quebec, Montreal, QC, Canada. He is leading a research group that investigates smartphone, embedded systems, and IoT security. His research interests include cloud security and secure sharing of embedded systems.}

\authorbibliography[scale=0.35, wraplines=7, overhang=40pt, imagewidth=0.25\textwidth, imagepos=r]
{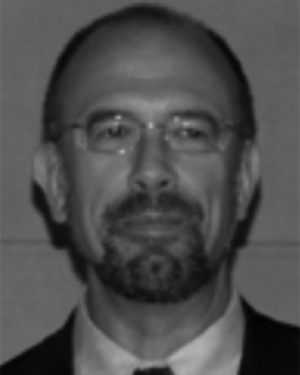}{Ernesto Damiani}{
is currently a Full Professor at the Department of Computer Science, Universita degli Studi di Milano, where he leads the Secure Service-oriented Architectures Research (SESAR) Laboratory. He is also the Founding Director of the Center for Cyber–Physical Systems, Khalifa University, United Arab Emirates. He received an Honorary Doctorate from Institute National des Sciences Appliquees de Lyon, France, in 2017, for his contributions to research and teaching on big data analytics. He is the Principal Investigator of the H2020 TOREADOR project on Big Data as a Service. He serves as Editor in Chief for IEEE Transactions on Services Computing. His research interests include cyber-security, big data, and cloud/edge processing, and he has published over 600 peer-reviewed articles and books. He is a Distinguished Scientist of ACM and was a recipient of the 2017 Stephen Yau Award.}

\end{document}